\documentstyle[12pt]{article}
\let\section=\subsection     \let\subsection=\subsubsection                
\newcommand{\spc}{{\ }}
\newcommand{\pr}[1]{{\sc{\lowercase{#1}}}}

\newcommand{\Tb}{$^{151}$Tb}
\newcommand{\Dy}{$^{152}$Dy}

\newcommand{\Pb}{$^{208}$Pb}

\newcommand{\codeversion}{1.60n}

\newcommand{\DY}{$^{152}_{~66}$Dy$_{86}$}

\newcounter{leteq}
\newcommand{\steplet}{\stepcounter{leteq}\addtocounter{equation}{-1}}

\newenvironment{eqnalpha}{\setcounter{leteq}{1}

\begin{eqnarray}}{\end{eqnarray}%
}

\newenvironment{eqnalphalabel}[1]{\setcounter{leteq}{1}
\raisebox{0cm}[0cm][0cm]{\begin{minipage}{1cm}%
\begin{eqnarray}\label{#1}&&\nonumber\end{eqnarray}\end{minipage}}

\begin{eqnarray}}{\end{eqnarray}%
}

\newcommand{\bnl}{\begin{eqnalpha}}
\newcommand{\enl}{\end{eqnalpha}}
\newcommand{\bnll}[1]{\begin{eqnalphalabel}{#1}}
\newcommand{\enll}{\end{eqnalphalabel}}

\newcommand{\text}[1]{\mbox{\scriptsize{#1}}}

\newcommand{\key}[1]{\vspace{1ex}\noindent{\it Keyword:}{\spc}{\tt{#1}}
                         \newline\phantom{{\it Keyword:}{\spc}{\tt{#1}}}{\spc}}

\newcommand{\be}{\begin{equation}}
\newcommand{\ee}{\end{equation}}
\newcommand{\ba}{\begin{array}}
\newcommand{\ea}{\end{array}}
\newcommand{\bn}{\begin{eqnarray}}
\newcommand{\en}{\end{eqnarray}}
\newcommand{\bc}{\begin{center}}
\newcommand{\ec}{\end{center}}
\newcommand{\bi}{\begin{itemize}}
\newcommand{\ei}{\end{itemize}}

\newcommand{\tfrac}[2]{{\textstyle{\frac{#1}{#2}}}}
\newcommand{\thalf}{{\textstyle{\frac{1}{2}}}}

\newcommand{\jde}{{{\cal J}^{(2)}}}
\newcommand{\jun}{{{\cal J}^{(1)}}}
\newcommand{\umi}{{$\hbar^2$/MeV}}
\newcommand{\ho}{{\hbar\omega}}
\newcommand{\hozero}{{\hbar\omega_0}}

\newcommand{\hox}{{\hbar\omega_x}}
\newcommand{\hoy}{{\hbar\omega_y}}
\newcommand{\hoz}{{\hbar\omega_z}}

\begin{document}

\hfill{CRN 96-31}

\vspace{0.5cm}
\begin{center}
        {\bf\Large
                     Solution of the Skyrme-Hartree-Fock equations in
                     the Cartesian deformed harmonic oscillator basis.
                            (II) The program \pr{HFODD}.
        }

\vspace{5mm}
        {\large
                       J.Dobaczewski$^{a,b,}$\footnote
                       {E-mail: dobaczew@fuw.edu.pl}
                   and J. Dudek$^{a,}$\footnote
                       {E-mail: jerzy@crnhp1.in2p3.fr}
        }

\vspace{3mm}
        {\it
              $^{a}$Centre de Recherches Nucl\'eaires,
                    IN$_2$P$_3$--CNRS/Universit\'e Louis Pasteur, Strasbourg I\\
                    F-67037 Strasbourg Cedex 2, France                        \\
              $^{b}$Institute of Theoretical Physics, Warsaw University       \\
                      ul. Ho\.za 69, PL-00681 Warsaw, Poland
        }
\end{center}

\vspace{5mm}
\hrule

\vspace{2mm}
\noindent{\bf Abstract}

We describe the code \pr{HFODD} which solves the nuclear
Skyrme-Hartree-Fock problem by using the deformed Cartesian
harmonic oscillator basis. The user has a possibility
of choosing among various symmetries of the nuclear HF problem for rotating
or non-rotating
nuclei; they vary from the non-axial parity-invariant
nuclear shapes, through those also breaking the
intrinsic parity, towards the least-restrictive case corresponding
to only one symmetry plane. The code provides a solution for a
complete superdeformed rotational band in an $A$$\sim$150 nucleus
within one CPU hour of the CRAY C-90 supercomputer or within
two-three CPU hours of a fast workstation.

\vspace{2mm}
\hrule

\vspace{2mm}


\vspace{5mm}
{\bf\large PROGRAM SUMMARY}

\bigskip\noindent{\it Title of the program:} \pr{HFODD}
                 (v\codeversion)

\bigskip\noindent{\it Catalogue number:}

\bigskip\noindent{\it Program obtainable from:}
                      CPC Program Library, \
                      Queen's University of Belfast, N. Ireland
                      (see application form in this issue)

\bigskip\noindent{\it Licensing provisions:} none

\bigskip\noindent{\it Computer for which the program is designed and
                      others on which it has been tested:}
                      CRAY C-90, SG Power Challenge L, IBM RS/6000

\bigskip\noindent{\it Operating systems:} UNIX, UNICOS, IRIX, AIX

\bigskip\noindent{\it Programming language used:} FORTRAN-77

\bigskip\noindent{\it Memory required to execute with typical data:} 10 Mwords

\bigskip\noindent{\it No. of bits in a word:} 64

\bigskip\noindent{\it Has the code been vectorised?:} Yes

\bigskip\noindent{\it No.{\spc}of lines in distributed program:}
                      19 438 (of which 8 354 are comments and separators)


\bigskip\noindent{\it Nature of physical problem}

\noindent
The nuclear mean-field and an analysis of its symmetries in realistic cases
are the main
ingredients of a description of nuclear states. Within the
Local Density Approximation, or for a zero-range velocity-dependent
Skyrme interaction, the nuclear mean-field is local and velocity
dependent. This allows an effective and fast solution of the
self-consistent Hartree-Fock
equations even for heavy nuclei and for different
configurations, deformations, excitation energies, or angular momenta.

\bigskip\noindent{\it Method of solution}

\noindent
The program uses the Cartesian harmonic oscillator basis to
expand single-particle wave functions of neutrons and protons
interacting by means of the Skyrme effective interaction.  The expansion
coefficients are determined by the iterative diagonalization of
the mean field Hamiltonians or Routhians which depend nonlinearly
on the local neutron and proton densities.  Suitable constraints
are used to obtain states corresponding to a given configuration,
deformation or angular momentum.

\bigskip\noindent{\it Restrictions on the complexity of the problem}

\noindent
The main restriction is the CPU time required for calculations
of heavy deformed nuclei and for a given precision required.
One symmetry plane is assumed. Pairing correlations are only
included in the BCS limit and for the conserved time-reversal symmetry.

\bigskip\noindent{\it Typical running time}

\noindent
One Hartree-Fock iteration for the superdeformed, rotating,
parity conserving state of {\DY} takes about
nine seconds on the CRAY C-90 computer.  Starting from the
Woods-Saxon wave functions, about fifty iterations are required
to obtain the energy converged within the precision of about
0.1\,keV.  In case when every value of the angular velocity is
converged separately, the complete superdeformed band with
precisely determined dynamical moments $\jde$ can be obtained
within one hour of CPU on the CRAY C-90, or within two to three
hours of CPU on the SG Power Challenge L or IBM RS/6000
computers. This time can be often reduced by a factor of three
when a self-consistent solution for a given rotational frequency is used
as a starting point for a neighboring rotational frequency.

\bigskip\noindent{\it Unusual features of the program}

\noindent
The user must have an access to the NAGLIB subroutine \pr{F02AXE}
or to the ESSL subroutine \pr{ZHPEV} which
diagonalize complex hermitian matrices, or provide another
subroutine which can perform such a task.

\noindent
The code is written in
single-precision for the use on a 64-bit processor. The compiler
option {\tt{}-r8} or  {\tt{}+autodblpad} (or equivalent) has to be used
to promote all real and complex single-precision floating-point items
to double precision when the code is used on a 32-bit machine.


\bigskip

\bigskip

{\bf\large LONG WRITE UP}

\bigskip

\section{Introduction}
\label{sec1a}

The method of solving the Skyrme-Hartree-Fock (HF) equations in the
Cartesian harmonic oscillator (HO) basis was described in the
previous publication \cite{comcom1} which is referred to as I.
The present paper is a long write up of the code \pr{HFODD} which
implements this method and solves the three-dimensional HF
equations. A one-dimensional Skyrme-HF code
restricted to the spherical symmetry has been published
by P.-G.~Reinhard \cite{Rei91}. Although several HF codes allowing triaxial
deformations exist \cite{Gir83,Bon85,Uma91,Chi95}, the code \pr{HFODD}
is the first one of that kind which is made available in the public domain.
Several earlier beta versions of the code have already been
distributed and used by various groups.  The present version
(v\codeversion)  is published for the first time and replaces all
previous versions.

In Sec.{\spc}\ref{sec5a} we present the numerical tests of the
code, and in Secs.{\spc}\ref{sec6}, \ref{sec6j}, and \ref{sec6v}
we describe its input, output, and source files, respectively.

\section{Numerical tests}
\label{sec5a}

Accuracy of the solution of the HF equations with the wave
functions expanded onto the Cartesian HO basis,
Eq.{\spc}(I-\ref{eq544}),\footnote{Symbol (I-\ref{eq544}) refers
to Eq.{\spc}(\ref{eq544}) of I.}
depends on the three parameters $\hox$, $\hoy$, $\hoz$ defining
the HO frequencies in three Cartesian directions, and on the
number $M$ of the HO states included in the basis.  In the
code \pr{HFODD} we use the standard prescription \cite{Dam69,FQK73} to chose
the HO states included in the basis, namely, the $M$ states with
the lowest HO single-particle energies,
   \be\label{eq592}
       \epsilon_{n_xn_yn_z} = \hox(n_x+\thalf) +
                              \hoy(n_y+\thalf) +
                              \hoz(n_z+\thalf),
    \ee
are selected among those which have $n_x$$\leq$$N_0$,
$n_y$$\leq$$N_0$, and $n_z$$\leq$$N_0$, where $N_0$ is the fixed
maximum number of HO quanta.  It should be noted that in general
both $M$ and $N_0$ have to be specified to define the basis.
Only for large $N_0$, the basis is defined solely by $M$ and does
not depend on $N_0$.  In this case, the grid of points
($n_x,n_y,n_z$) defining the states included in the basis forms a
pyramid in three dimensions, with the inclined face delimited by the
condition $\epsilon_{n_xn_yn_z}$$\leq$const.  On the other hand,
only for small values of $N_0$ the basis is defined solely by
$N_0$ and does not depend on the energy cut-off.  In this case
the corresponding grid of points $n_xn_yn_z$ forms a cube of the
size $N_0$.  In all intermediate cases the shape of the basis
corresponds to a pyramid with the corners cut off, or to a cube
with the corners cut off.  Usually $N_0$ is chosen large enough so
that all the states allowed by the energy cut-off are included in the
basis.  The shape of the basis is printed in form of three
projections by the code \pr{HFODD} on the output file, see the
TEST RUN OUTPUT below.

In what follows we discuss only the results obtained with axially
deformed bases for $\omega_\perp$=$\omega_x$=$\omega_y$, and present the
results in terms of the standard parameters
   \be\label{eq601}
      \omega_0=(\omega_\perp^2\omega_z)^{1/3}\quad \mbox{and} \quad
      q=\omega_\perp/\omega_z .
    \ee
(The code accepts of course in general a triaxially deformed HO basis
what offers the user some additional flexibility in approaching the
actual physical problem).
In Ref.{\spc}\cite{FQK73} the dependence of results on the basis
parameters and on the basis size has been tested for
$N_0$$\leq$10, which at that time was at the limit of the numerical
feasibility.  In the present study we discuss
similar tests up to $N_0$=20 (spherical case) and $N_0$=26
(deformed case).

{}From the tests based on low values of $N_0$, the authors of
Ref.{\spc}\cite{FQK73} concluded that the basis parameters
$\omega_0$ and $q$ have to be optimized for a given basis size in
such a way that the HF energy is minimized.  This procedure is
consistent with the variational character of the HF theory in
which the best approximation to the energy is obtained by a
minimization of the energy.  Although the variational principle
does not ensure that the best approximation to other observables
is obtained for the wave functions corresponding to the minimum
of energy, this turned out to be the case for low values of
$N_0$.  The necessity to optimize the basis parameters was the
real bottleneck of the method, because it requires many HF
calculations to be performed before one physical solution can
been obtained.

The picture described above does not apply to calculations performed
with relatively large values of $N_0$.  Of course, the best
energy is still obtained with optimized values of the basis
parameters, but the energy depends now very weakly on the basis
parameters and the error generated by using a non-optimal basis
is small.  Moreover, the best results for other observables are
not necessarily obtained with the optimal basis parameters.  This
can be well understood by noticing that the minimum obtained from
a very weak dependence of the energy on the basis parameters may
be illconditioned, and may occur at rather accidental value.
Therefore, for large values of $N_0$, the basis optimization does
not bring any substantial improvement of precision.  Instead of
the optimization, suitable ``physical'' values of the basis
parameters can be determined based on simple geometrical
considerations.  This is in line with results of
Ref.{\spc}\cite{FQK73}, where it was shown that a partial
optimization followed by the liquid-drop-model estimates
reproduces the optimal basis parameters even for small values of
$N_0$.

In the code \pr{HFODD} the physical basis parameters
are determined as follows (cf.{\spc}also Ref.{\spc}\cite{Gir83}).
First we define $\omega_0$ by the standard value \cite{SGN69}
multiplied by some phenomenological factor $f$ close to unity,
   \be\label{eq594}
   \hozero=f\times41\,\mbox{MeV}/A^{1/3}.
   \ee
Based on the experience gained from the diagonalization of
the Woods-Saxon Hamiltonian on the HO basis \cite{DST81,CDN87},
the recommended
value is $f$=1.2. This particular scaling of the oscillator
frequencies turns out
to help in stabilizing the results of the diagonalization with
respect to adding
the new basis states.
Second, suppose
we wish to perform the HF calculations for a given set of
multipole deformations $\alpha_{\lambda\mu}$, i.e., for a nuclear
shape defined by the surface $\Sigma$
   \begin{equation}\label{eq101}
     \Sigma:\,\,R(\theta,\phi)=c(\alpha)
         \left[1+\sum_{\lambda=0}^{\lambda_{max}}
         \sum_{\mu=-\lambda}^{\lambda}\alpha_{\lambda\mu}
         Y_{\lambda\mu}(\theta,\phi)\right],
   \end{equation}
where $c(\alpha)$ is a function of $\alpha_{\lambda\mu}$ such
that the volume enclosed by the surface $\Sigma$ does not depend
on $\alpha$. Due to the assumed $y$-simplex symmetry,
see Sec.{\spc}\ref{sec9a} of I, all
multipole deformations $\alpha_{\lambda\mu}$ are real, and only
those with $\mu$$\geq$0 are used in the code \pr{HFODD} because
then $\alpha_{\lambda,-\mu}$=$(-1)^{\mu}\alpha_{\lambda\mu}$.

The lengths of principal axes of the volume enclosed in the
surface $\Sigma$ can be defined
as $R_x$=$R(\pi/2,0)$, $R_y$=$R(\pi/2,\pi/2)$, and $R_z$=$R(0,0)$.
For complicated shapes they do not necessarily correspond to
the principal axes of the moment of inertia or quadrupole moment.
However, when the quadrupole deformation dominates, this will
roughly be so if the quadrupole deformation is considered in
its intrinsic frame of reference, i.e., if one uses
$\alpha_{21}$=0.
The relative values of the HO frequencies are then defined by the
condition
   \be\label{eq593}
   \omega_x{}R_x=\omega_y{}R_y=\omega_z{}R_z,
   \ee
while the common proportionality constant is given by the condition
$(\omega_x\omega_y\omega_z)^{1/3}$=$\omega_0$.

The above prescription is very helpful if the values of physical
deformations are (at least roughly) known for a given nucleus.
This is true in particular for the
superdeformed shapes which have very similar deformations in a
given region of nuclei.  For example, the superdeformed nuclei in
the $A$$\sim$150 region can be very well described in the basis
corresponding to the {\Dy} nucleus with deformations
$\alpha_{20}$=0.61 and $\alpha_{40}$=0.10 \cite{J2D2b}.  This
gives the physical basis parameters, $\hox$=$\hoy$=6.246\,MeV and
$\hoz$=11.200\,MeV, i.e., $\hozero$=9.219\,MeV and $q$=1.793.

We may now present results of tests performed for
different basis parameters and compare them with those obtained
for the physical basis parameters.  All calculations below have
been performed for the Skyrme parametrization SkM* with the
coupling constants $C_t^{J}$=$C_t^{T}$=$C_t^{\Delta s}$=0
in the energy functional (I-\ref{eq109}).  The
latter conditions correspond to simplifications of the Skyrme
functional usually assumed in studies made within
the spatial coordinates \cite{BON87a,BON91a}, and allow a
comparison of our results with those obtained by an independent
method and an independent HF code.

{}From now on we are going to use the two adjectives introduced above and
associated with the
basis parameters: {\em optimal}, i.e., adjusted to a given basis size so as
to minimize the HF energy, and {\em physical}, i.e.,
corresponding to the actual
information about the deformations of nuclei studied, and calculated
through Eqs.{\spc}(\ref{eq594}), (\ref{eq101}), and (\ref{eq593}) based on the
experimental result or earlier Strutinsky-type calculations.

\subsection{Spherical nuclei}
\label{sec5b}

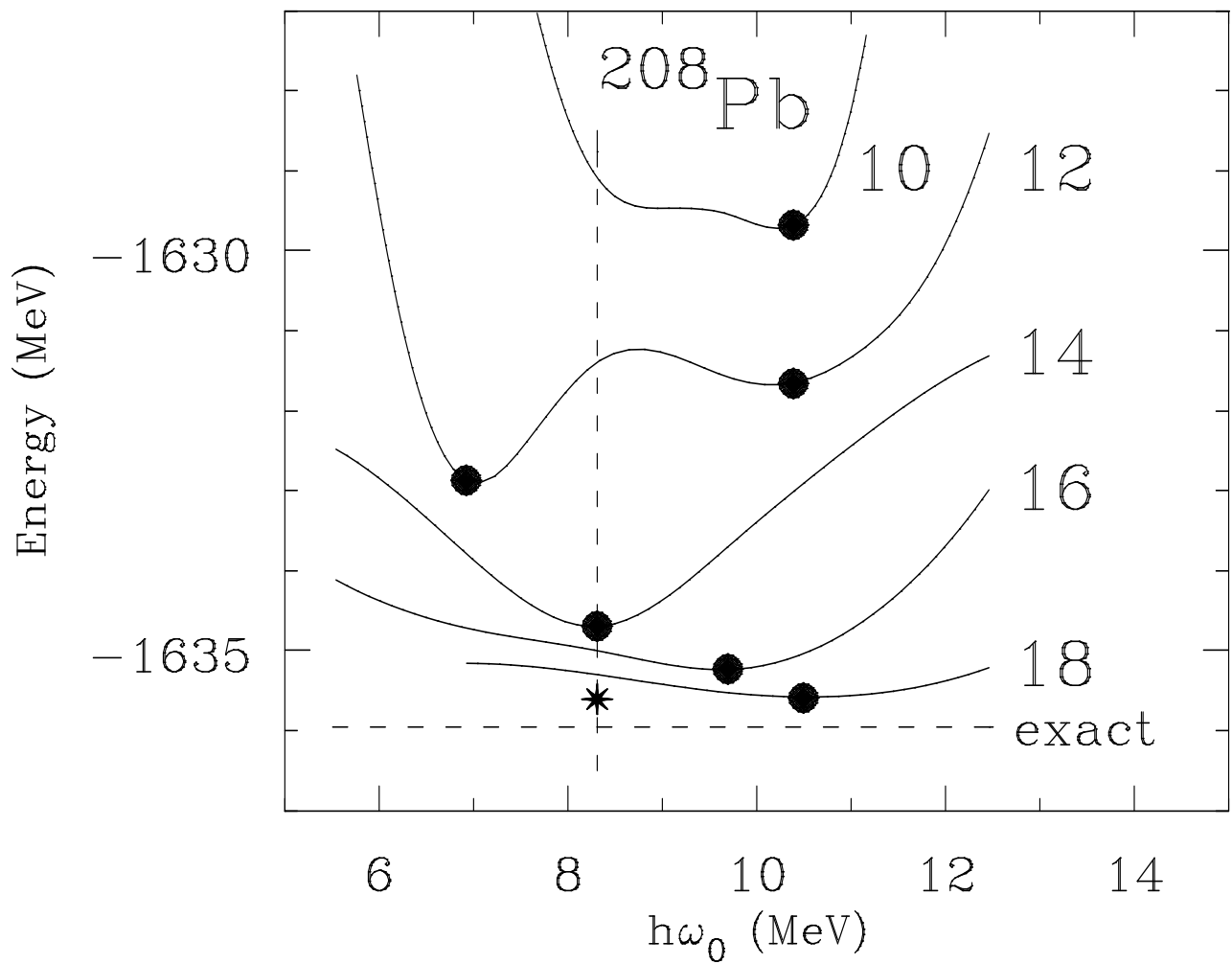
\begin{figure}[ht]
\caption[F1]{%
Ground-state energies of {\Pb} (solid lines)
calculated as functions of $\hozero$ for several values of the number
$N_0$ of
HO shells included in the basis. Dots denote the minima,
the asterisk gives the value calculated at $N_0$=20 for the
physical value of $\hozero$ shown by the vertical dashed line.
The horizontal dashed line represents the exact result.
}
\label{fig01}
\end{figure}

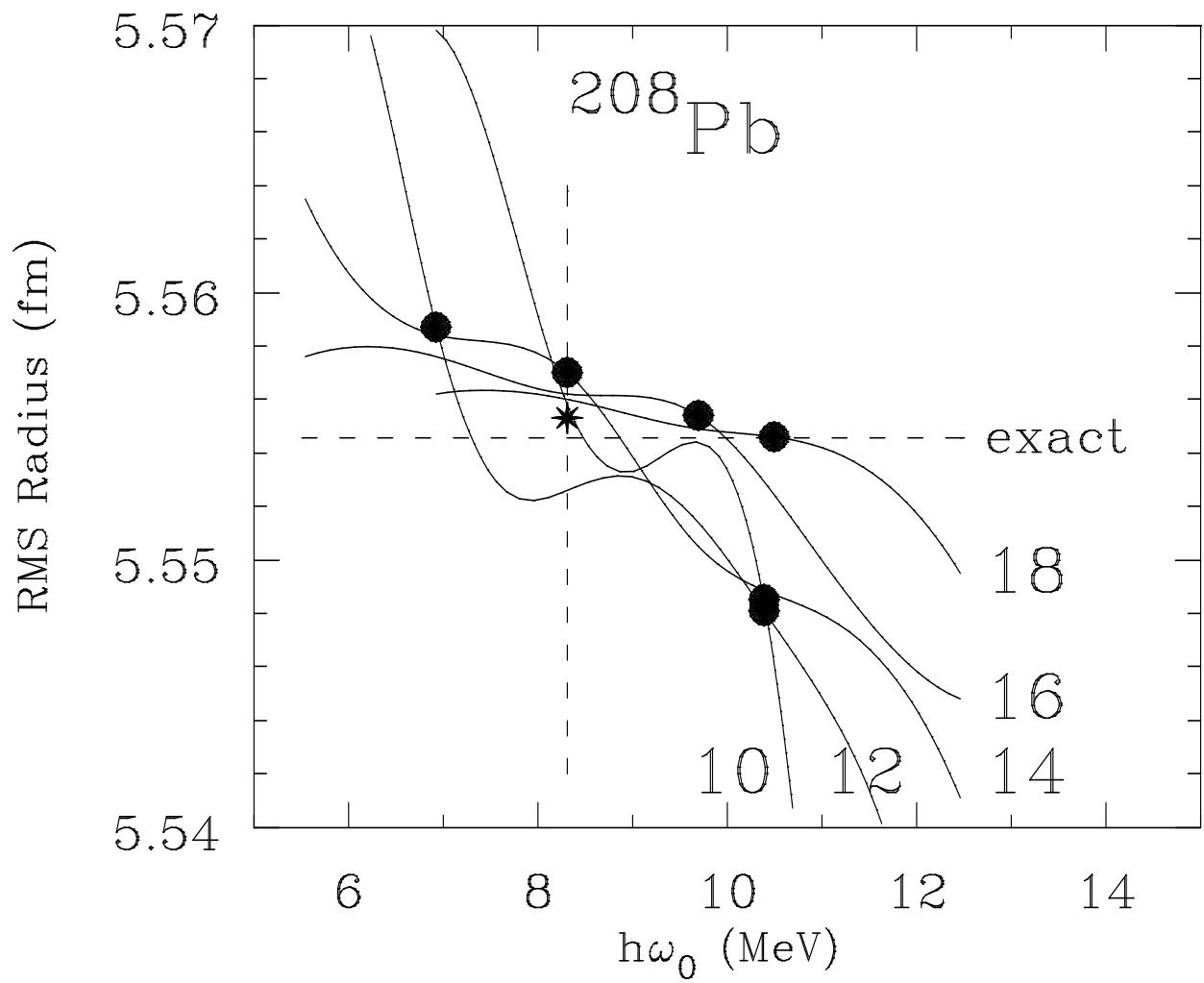
\begin{figure}[ht]
\caption[F2]{%
Same as in Fig.{\spc}\protect\ref{fig01} but for the rms radii
of {\Pb}.
}
\label{fig02}
\end{figure}

We begin with the results obtained for a spherical nucleus {\Pb}
for which the spherical basis is used, $q$=1, and the results are
studied in function of $\hozero$.  The physical value for this
nucleus is $\hozero$=8.304\,MeV, and is represented in
Figs.{\spc}\ref{fig01} and \ref{fig02} by the vertical dashed
lines.  The horizontal dashed lines represent very precise
solutions obtained by the simple one-dimensional spherical code
in spatial coordinates, and are considered as the exact values.
The solid lines represent the results calculated by the code
\pr{HFODD} for $N_0$ from 10 to 18, and the asterisk represents
the calculation with $N_0$=20.  For the spherical bases the
number of the HO states is given by $M$=($N_0$+1)($N_0$+2)($N_0$+3)/6,
and hence in Figs.{\spc}\ref{fig01} and \ref{fig02} it increases from
286 to 1771.

The energies of {\Pb}, shown in Fig.{\spc}\ref{fig01}, converge
rather slowly to the exact value of $-$1635.956\,MeV.  For
$N_0$=10 the error is about 6\,MeV and only for $N_0$$\geq$16
becomes smaller than 1\,MeV.  For every value of $N_0$, the dot
denotes the minimum of energy which defines the optimal value
of $\hozero$.  For $N_0$=10 and 12 the optimal values
are rather remote from the physical value and,
moreover, two minima appear for $N_0$=12.  For $N_0$=14 the
minimum occurs at the physical value of $\hozero$, but for larger
bases, $N_0$=16 and 18, the minimum becomes poorly defined and
deviates from the physical value of $\hozero$.

{}Fig.{\spc}\ref{fig02} shows the corresponding results of
calculations for the matter rms radii of {\Pb}.  For every curve
the dots represent the minima of the energies shown in
Fig.{\spc}\ref{fig01}.  One should note that the vertical scale
of the Figure is very much expanded and shows a narrow region of
radii within about 0.3\% around the exact value of
$R_{\text{rms}}$=5.5546\,fm.  For $N_0$=10 the radii significantly
depend on $\hozero$ but the value of $R_{\text{rms}}$ at the physical
basis parameter is very close to the exact result, while the
value at the optimal basis parameter is much worse.  For
$N_0$=12 the value at the physical $\hozero$ is again much better
than those at any of the two optimal bases.  For $N_0$=14 the
physical and optimal bases coincide and give the exact value with
the error of 0.05\%, while for larger bases the dependence of
$R_{\text{rms}}$ on $\hozero$ becomes weak, and the results for the
optimal bases are only slightly better than those for the
physical basis.  Altogether, for any value of $N_0$$\geq$10 the
physical basis gives the radii precise up to $\pm$0.003\,fm, or
up to $\pm$0.05\%.

\subsection{Deformed nuclei}
\label{sec5c}

In order to study the properties of the optimization of the
deformation of the basis, we have performed a series of
calculations for the non-rotating superdeformed state in {\Dy}
($\ho$=0)
with $\hozero$ fixed at the physical value, and for several different
values of $q$ and $M$.  These results can be compared to those obtained
in spatial coordinates for the same nucleus and force, and with
the three-dimensional cubic grid of points spaced by
$\Delta{x}$=1.0, 0.8, or 0.7\,fm.  In this case smaller spacing
gives more accurate results, and one obtains \cite{PH},
respectively, the total energies ${\cal E}$=$-$1234.611,
$-$1230.769, or $-$1230.104\,MeV and the proton quadrupole
moments $Q_p$=18.316, 18.369, or 18.376\,b.

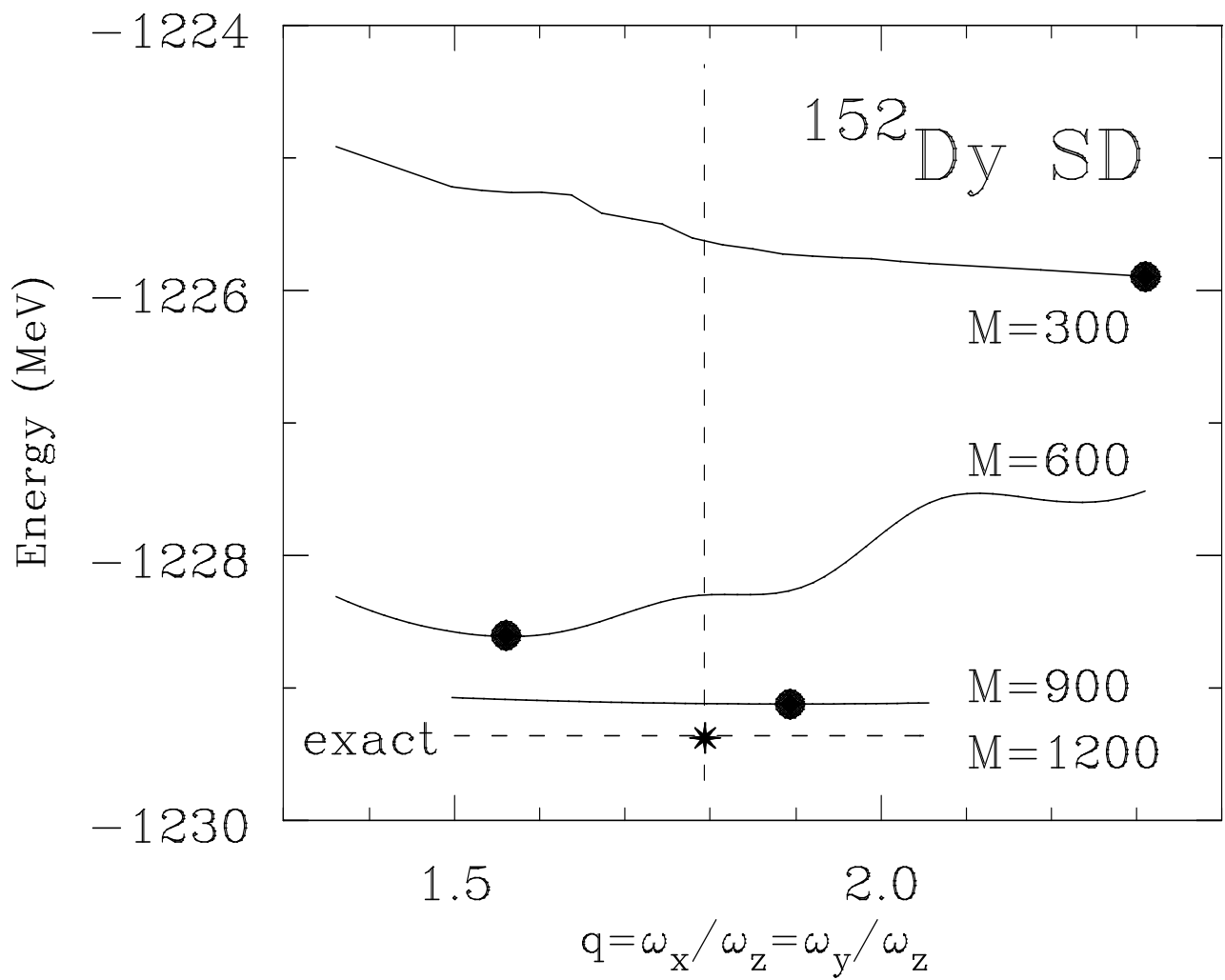
\begin{figure}[ht]
\caption[F3]{%
Energies of superdeformed state in {\Dy} (solid lines)
calculated at $\ho$=0
as functions of the basis deformation parameter $q$
for several values of the number $M$ of states included in the
HO basis.
Dots denote the minima,
the asterisk gives the value calculated at $M$=1200 for the
physical value of $q$ shown by the vertical dashed line.
The horizontal dashed line represents the exact result,
see text.
}
\label{fig03}
\end{figure}

These values of energies were obtained by using the finite
difference expressions for derivatives, which are less precise
than the Fourier expressions \cite{Bay86}.  If the HF equations
are solved using the first type of approximation but nevertheless
at the end of the convergence the energies are calculated with the
second method, one obtains a good and cost-effective estimate of
the energy \cite{Bay86,Hee91}.  Such a procedure yields \cite{PH}
for $\Delta{x}$=0.7\,fm the value of ${\cal E}$=$-$1229.365 which we
may consider for the purpose of the discussion below as the exact
result.  This energy is shown by the horizontal dashed line in
Fig.{\spc}\ref{fig03}.  In practice, the three-dimensional
calculations in the spatial coordinates are usually done with
$\Delta{x}$=1\,fm and employ the finite difference method.  For
these standard conditions the binding energy of {\Dy} is overestimated by
about 5\,MeV.

Fig.{\spc}\ref{fig03} shows the total energies in {\Dy}
calculated by the code \pr{HFODD} for $M$=300, 600, 900, or 1200.
When $q$ varies from 1.3 to 2.3 the maximum numbers of the HO
quanta in the directions perpendicular/parallel to the elongation
axis vary for $M$=300 from 9/13 to 8/18 quanta.
Similarly, for $M$=1200 and $q$=1.793 the HO basis contains
14/26 quanta.
Here and in the following, the results
for $M$=300 are presented as broken lines with visible jumps
corresponding to individual states entering and leaving the HO
basis.  Indeed, keeping a fixed number $M$ of states with varying
$q$ means that a given HO orbital may cross the boundary
$\epsilon_{n_xn_yn_z}$$\leq$const.  For relatively small values of $M$ this
may create some effects on various quantities (visible in the
expanded scales of presented Figures), while for larger values of $M$
such effects disappear.

The convergence of energy in function of $M$ is again
rather slow; for $M$=300 the error is around 5\,MeV and it
decreases to below 1\,MeV only at $M$$>$600.  For $M$=300 and 600
the optimal basis deformations are rather different from the
physical value.  However, the energy gain from optimizing the
value of $q$ is very small.  For $M$=900 the optimal and physical
values of $q$ are close to one another, but the minimum of energy is hardly
visible.  For $M$=1200 we obtain ${\cal E}$=$-$1229.383 in a very
good agreement with the exact result. I should be noted that
both estimates compared here are variational, and hence both may still
slightly decrease with further improvement of the numerical parameters.

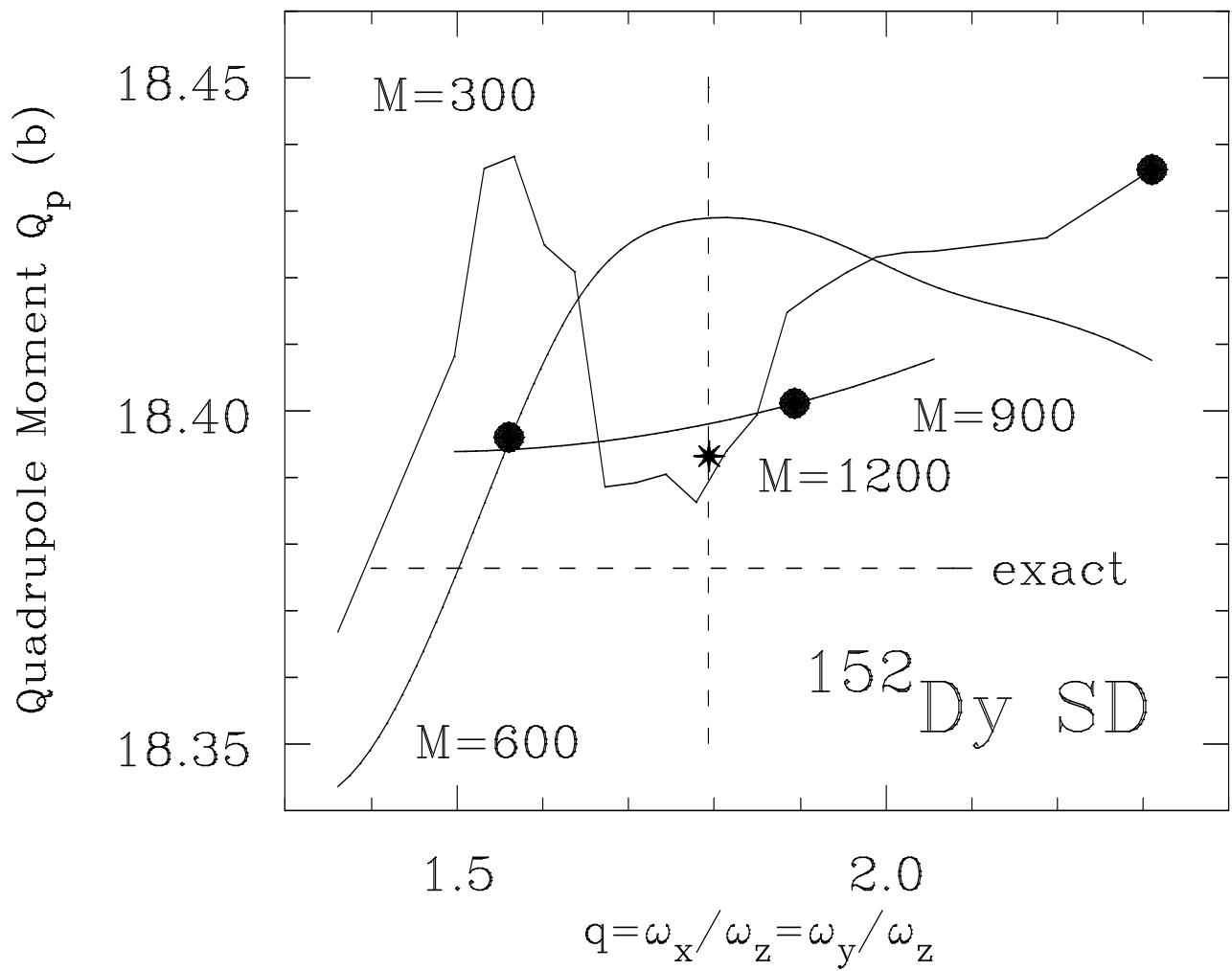
\begin{figure}[ht]
\caption[F4]{%
Same as in Fig.{\spc}\protect\ref{fig03} but for the proton quadrupole
moments $Q_p$ of superdeformed {\Dy} at $\ho$=0.
}
\label{fig04}
\end{figure}

{}Fig.{\spc}\ref{fig04} shows the values of proton quadrupole
moments calculated by the code \pr{HFODD} and compared with the
value of 18.376\,b obtained for $\Delta{x}$=0.7\,fm (see above).
The latter value is for the purpose of the present test
considered as the exact result.
Again the region shown on the ordinate is very narrow; within
0.5\% of the value of $Q_p$.  On the other hand, the region
of $q$ presented on the abscissa is fairly large; it corresponds
to the standard quadrupole deformations $\beta$$\equiv$$\alpha_{20}$ of the
nuclear surface (\ref{eq101}) ranging from 0.3 to 1.0.

{}For $M$=300 one obtains much a better value of $Q_p$ for the
physical basis than for the optimal basis.  This situation is
reversed for $M$=600.  For $M$=900 the dependence on basis
parameters is very weak and the results are very close to the
physical-basis value of $Q_p$=18.393 obtained for $M$=1200.
The values of $Q_p$ calculated by the code \pr{HFODD} tend to be
slightly larger than the exact value obtained within the
spatial-coordinate method. We have checked that the optimization
of $\hozero$ does not remove this residual difference.

Similarly as for tests performed in spherical nuclei, one can
conclude that the optimization of the basis does not, in general,
improve the results as compared to those obtained for the
physical basis.  Within a very limited space of $M$=300 HO states
one may obtain a very reasonable estimate of $Q_p$ when the
physical basis is used.  Taking into account the fact that the
discrepancy is larger for $M$=600, we may attribute a conservative
error of 0.05\,b, or 0.3\% to the calculated values of the
quadrupole moments.  A precision of this order is perfectly
sufficient for all practical purposes, and can of course be
improved by using larger bases, instead of optimizing the basis
parameters.
In fact, the obtained precision is comparable to that of
the spatial-coordinate method when it is
used with the standard value of $\Delta{x}$=1\,fm (see above).

\subsection{Rotating nuclei}
\label{sec5d}

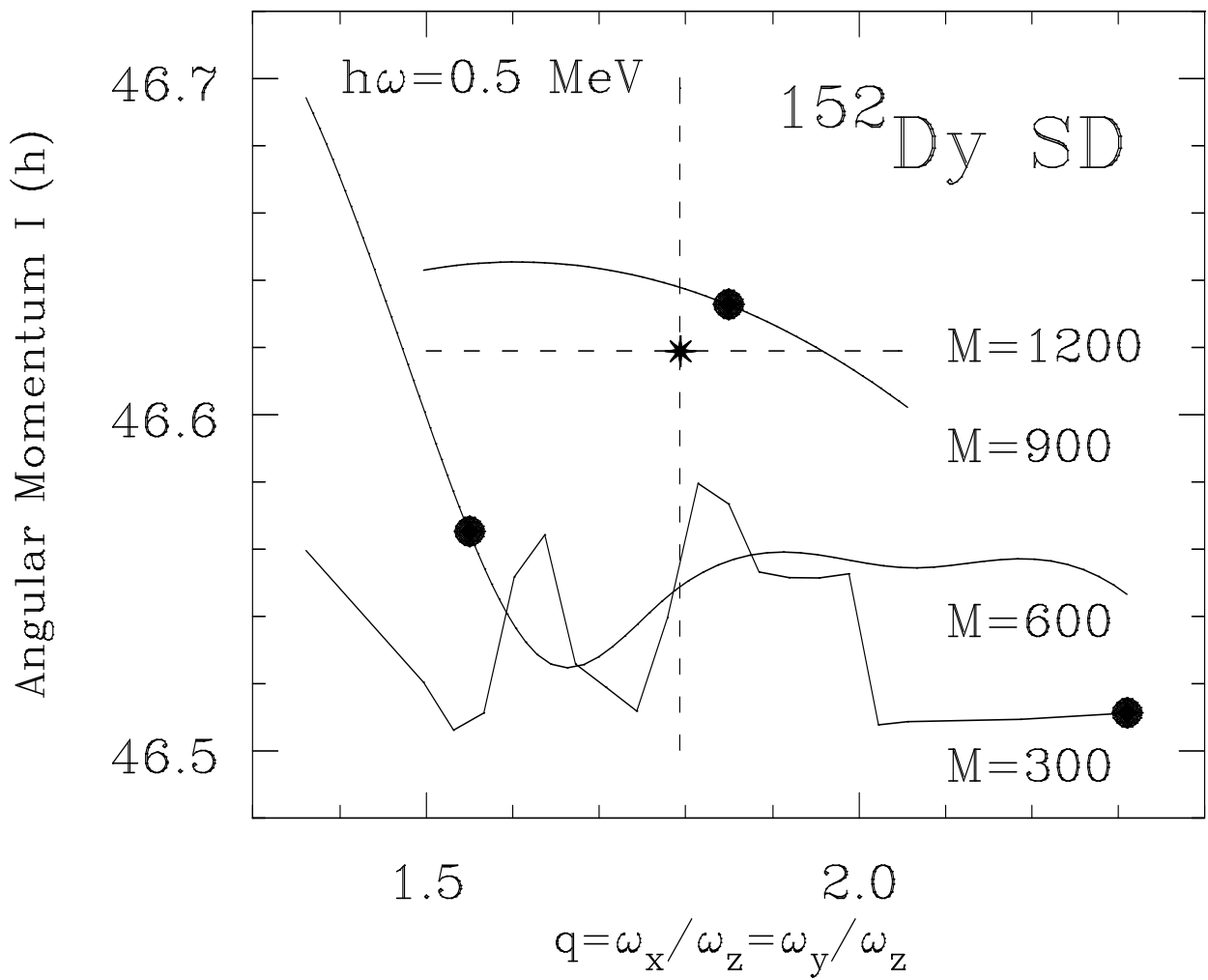
\begin{figure}[ht]
\caption[F4]{%
Angular momenta of the rotating superdeformed nucleus {\Dy} (solid lines)
calculated at $\ho$=0.5\,MeV
as functions of the basis deformation parameter $q$
for several values of the number $M$ of states included in the
HO basis.
Dots denote the minima of the Routhian (I-\ref{eq500}),
the asterisk and the horizontal dashed line
gives the value calculated at $M$=1200 for the
physical value of $q$ shown by the vertical dashed line.
}
\label{fig05}
\end{figure}

In Fig.{\spc}\ref{fig05} we show the values of the angular
momentum $I$ calculated for {\Dy} at $\ho$= 0.5\,MeV.  For
all values of $q$ and $M$ considered in the Figure one obtains
results which are within 0.1\,$\hbar$ of the value
calculated for $M$=1200.  Again, there is no advantage in using
the optimal bases as compared to the physical one.  The optimal
values of $q$ were obtained by minimizing the Routhian
(I-\ref{eq500}) at constant $\ho$.  We do not show here the
plot of the Routhian in function of $q$ because such a plot is very
similar to that of the energy at zero spin,
Fig.{\spc}\ref{fig03}, except for an almost constant shift by
$-$$\omega{I}$.

The properties of superdeformed bands are very often described
in terms of moments of inertia, which can be calculated as the
static or dynamic (first or second) moments
   \bnl
        \jun = \frac{I}{\omega} ,
                                               \label{eq603a} \\ \steplet
        \jde = \frac{d{I}}{d\omega} ,
                                               \label{eq603b}
   \enl
respectively.  The precision of determining $\jun$ is therefore
governed by that with which the total angular momentum $I$ is
calculated, and for $\ho$=0.5\,MeV it is equal to about
0.2\,\umi.  In practical calculations, the dynamic moment $\jde$
has to be calculated by the finite-difference expression
   \be\label{eq602}
   \jde = \frac{I(\omega_2)-I(\omega_1)}{\omega_2-\omega_1}
   \ee
which is precise up to
   \be\label{eq604}
   \delta\jde \sim \tfrac{1}{24}(\omega_2-\omega_1)^2
                           \frac{d^2\jde}{d\omega^2} .
   \ee
This expression for $\delta\jde$ corresponds to the lowest-order
term neglected in the Taylor expansion leading to the finite-difference
formula (\ref{eq602}).
The curvature of $\jde$, which determines its precision
$\delta\jde$, is usually fairly small (200--300\,$\hbar^4$/MeV$^3$
at most), except at the crossing
frequencies where the precise values of $\jde$ are
physically not important anyhow.  Therefore, the finite difference
expression (\ref{eq602}) can usually be used already at
$\omega_2$$-$$\omega_1$$\simeq$0.1\,MeV/$\hbar$.

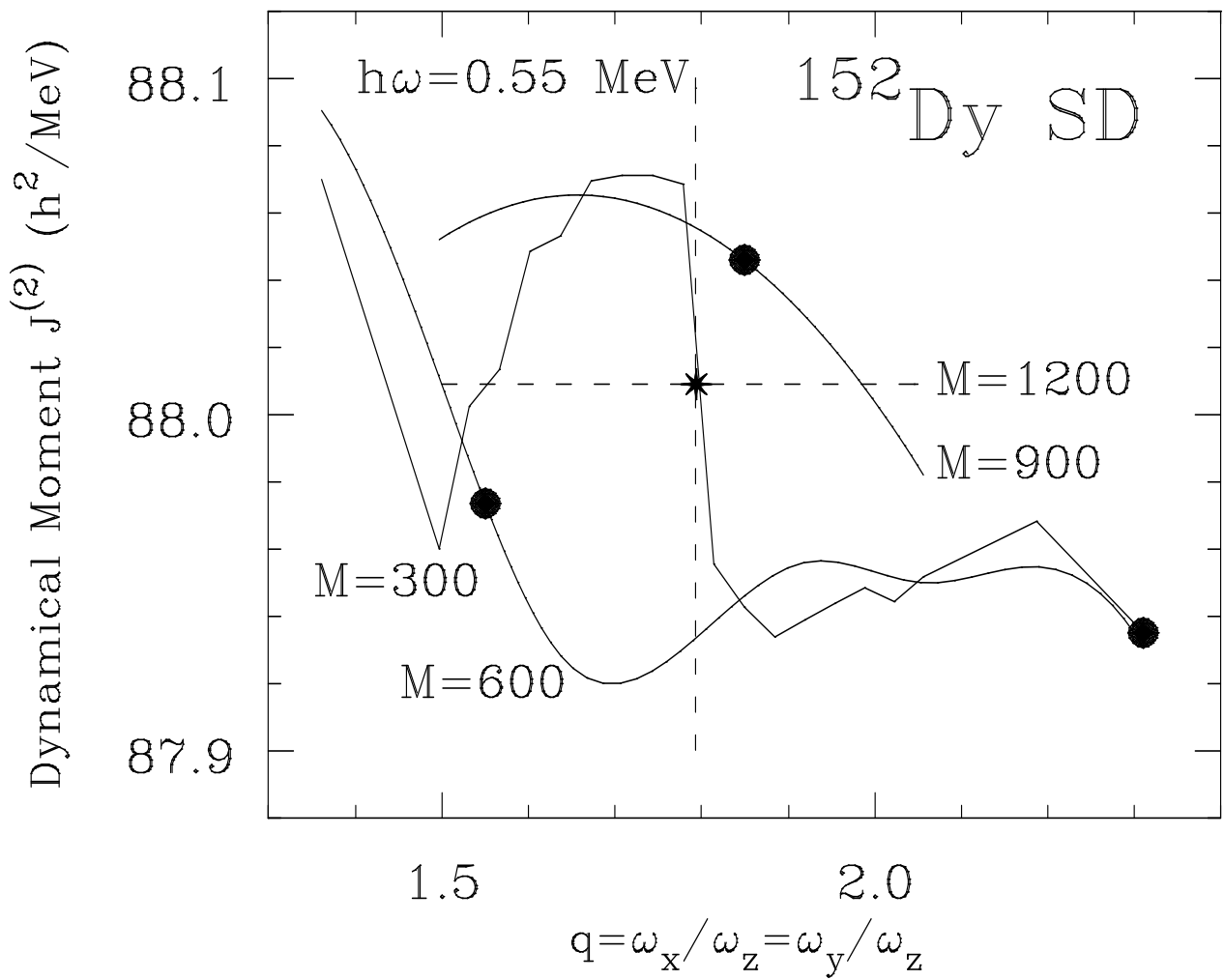
\begin{figure}[ht]
\caption[F4]{%
Same as in Fig.{\spc}\protect\ref{fig05} but for the
dynamic moment $\jde$ (\ref{eq602})
of superdeformed {\Dy} at $\ho$=0.55\,MeV.
}
\label{fig06}
\end{figure}

In Fig.{\spc}\ref{fig06} we present results for $\jde$ calculated at
$\ho$=0.55\,MeV by using the values of $I$ obtained at
$\omega_1$=0.5 and $\omega_2$=0.6\,MeV/$\hbar$.
In view of the precision of the
total angular momentum $I$, estimated above to be equal to 0.1\,$\hbar$,
one could have expected that the values of $\jde$ obtained in
this way will be precise only up to about 1\,\umi.  In fact, the
real precision of $\jde$ turns out to be an order of magnitude
{\em better}.  This is so because most of the errors related to the use
of the finite HO basis cancel out, and the difference of angular
momenta turns out to be much more precise that the total value of
$I$ itself.  As a result, the dynamic moments $\jde$ shown in
Fig.{\spc}\ref{fig06} are precise up to 0.1\,\umi.  This
observation concerns also many other observables, and will be discussed
in the next section in more detail.

\subsection{Relative energies, moments, and alignments}
\label{sec5e}

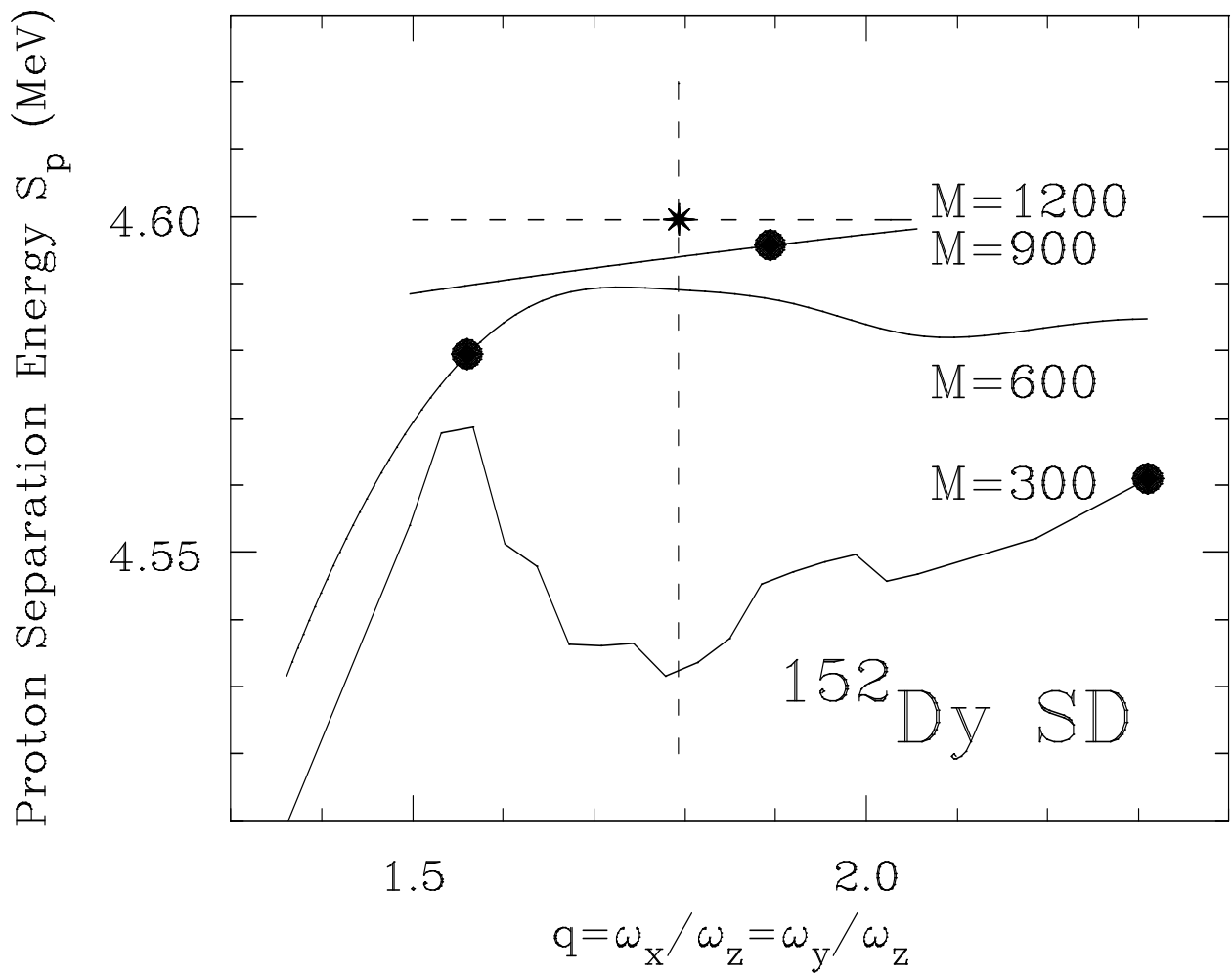
\begin{figure}[ht]
\caption[F4]{%
Proton separation energies in SD {\Dy} (solid lines)
relative to the SD state in {\Tb},
i.e., the differences of energies
${\cal E}^{\text{SD}}$(\Tb)$-$${\cal E}^{\text{SD}}$(\Dy),
obtained at $\ho$=0. Solid lines denote values calculated
as functions of the basis deformation parameter $q$
for several values of the number $M$ of states included in the
HO basis.
Dots denote the minima of the energy in {\Dy}
(Fig.{\spc}\protect\ref{fig03}),
the asterisk and the horizontal dashed line
gives the value calculated at $M$=1200 for the
physical value of $q$ shown by the vertical dashed line.
}
\label{fig07}
\end{figure}

As already pointed out in Ref.{\spc}\cite{FQK73}, the differences
of energies (the relative energies) depend much less on the basis
parameters than the total (absolute) energies.  This is illustrated in
Fig.{\spc}\ref{fig07}, where we show the proton separation
energies $S_p$ in the SD {\Dy} nucleus, calculated at spin zero for different
values of $q$ and $M$.  The values of $S_p$ have been calculated
by subtracting the total energy of {\Dy} from the total energy
of {\Tb} after both have been determined self-consistently by
using the same HO basis.  The SD state in {\Tb} has been
constructed by creating a hole in the $\pi$[651]3/2(+$i$)
orbital, cf.{\spc}Ref.{\spc}\cite{J2D2b}.  As seen in the Figure,
the separation energies are precise up to about 50\,keV, which is
two orders of magnitude better than the total energies in either
of {\Tb} or {\Dy} nucleus.  Similar precision is also expected
for the excitation energies in a given nucleus, i.e., for the
differences of energies corresponding to different configurations
or angular momenta.

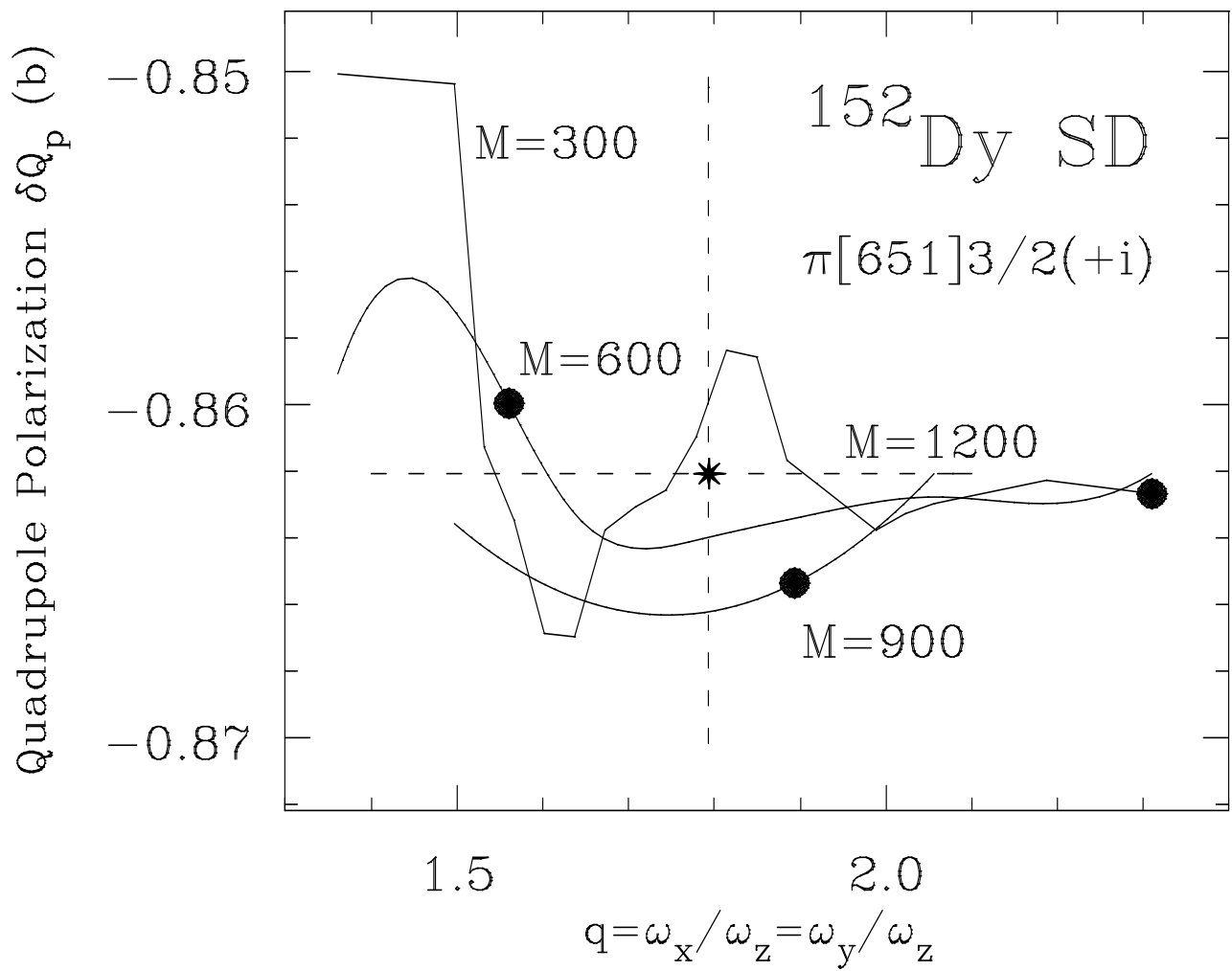
\begin{figure}[ht]
\caption[F4]{%
Same as in Fig.{\spc}\protect\ref{fig07} but for the
quadrupole polarizations induced in the {\Dy} core
by the $\pi$[651](3/2)(+$i$) hole,
i.e., the differences of proton quadrupole moments
$Q_p^{\text{SD}}$(\Tb)$-$$Q_p^{\text{SD}}$(\Dy).
}
\label{fig08}
\end{figure}

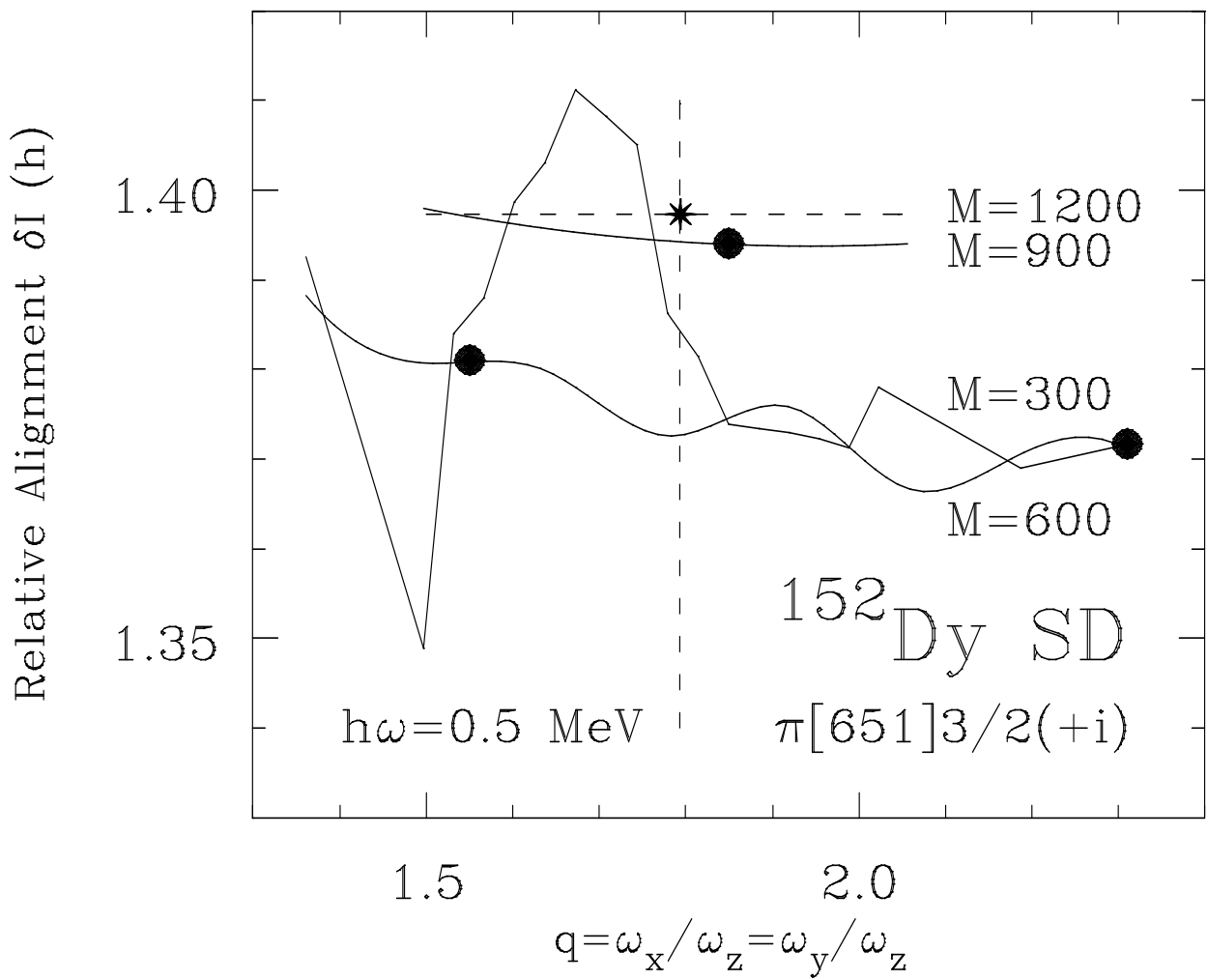
\begin{figure}[ht]
\caption[F4]{%
Same as in Fig.{\spc}\protect\ref{fig07} but for the
relative alignments between the {\Tb} and {\Dy} nuclei,
i.e., the differences of angular momenta
$I$(\Tb)$-$$I$(\Dy) at $\ho$=0.5\,MeV.
}
\label{fig09}
\end{figure}

The same principle also works for the quadrupole polarizations
\cite{Sat96} which are obtained by subtracting the quadrupole
moments of the SD states in {\Tb} and {\Dy} nuclei.  This is shown
in Fig.{\spc}\ref{fig08}, where the obtained quadrupole
polarizations are precise up to 0.01\,b, compared to the 0.05\,b
precision obtained for the values of proton quadrupole moments
$Q_p$ discussed above. Similarly, the relative alignments
between these two nuclei, i.e., the differences of angular momenta
at a given angular frequency, $\ho$=0.5\,MeV here,
are presented in Fig.{\spc}\ref{fig09}. These values characterize
the quality of the identical bands \cite{Bak95}, and have been
discussed in Ref.{\spc}\cite{J2D2b}. Here we see that using
the truncated HO basis they are precise up to 0.03\,$\hbar$,
which is a factor three better than the values of the
angular momenta discussed in Sec.{\spc}\ref{sec5d}.

\subsection{Summary}
\label{sec5f}

Table \ref{tab01} summarizes the results discussed in Sections
\ref{sec5b}--\ref{sec5e}.  The column denoted ``absolute values''
refers to the errors of observables calculated in {\Pb} or {\Dy}
for $M$=300 and for the physical basis parameters.  The errors are
deduced by comparing these results with the exact ones, where
available, or with the results obtained with $M$=1200.  The
percentage errors are also given.  The column denoted ``relative
values'' refers to the errors of differences of observables
calculated in {\Tb} and {\Dy}.  Due to the fact that most of the
errors are systematic (they originate from the truncated HO basis), they tend
to cancel out, and the errors of relative observables are smaller
than those of the absolute observables.  Since the relative
values are much smaller than the absolute values, the percentage
errors are larger in this case, but still only of the order of
several per cent.  In all cases the precisions of relative values
are sufficient for a meaningful comparison with data or with
other theories.  Whenever more precise results are required, they
can be easily obtained by increasing the size of the basis.  The
basis optimization is not a competing procedure which would lead
to a better precision.


\section{Input data file}
\label{sec6}

Different aspects of the organization of the code \pr{HFODD}
are presented in this section together with the description
of the input data file. The code reads the input data from
the standard FORTRAN input file, prints results on the
standard FORTRAN output file, and also writes two auxiliary
files which below are called the review file and the
record file. If required, the code may also start the
iteration from the Woods-Saxon wave functions, which have
to be provided in the specialized file called the Woods-Saxon file, or
from the previously recorded HF potentials provided in the file
called the replay file.

The general structure of the input data file is defined by
the following rules:
\begin{enumerate}
\item
Input data file is an ASCII file composed of independent items.
\item
An item is composed of the keyword line, and of the
data line which follows immediately the keyword line.
Two items contain no data lines, see point \ref{twoitem} below.
\item
The keyword line contains in columns 1 through 10 a keyword
which is a specific text defining the item. If the keyword
has less than 10 characters it has to be obligatorily padded
with appropriate number of trailing spaces. Columns beyond 10
are ignored and can be used to place arbitrary comments or texts.
\item
The data line contains a prescribed number of data values
which are read in the free FORTRAN FORMAT. The type
of data ({\tt{}REAL}, {\tt{}INTEGER}, or {\tt{}CHARACTER})
should match
the definition of the given data line.
The {\tt{}CHARACTER} data should be provided
starting from the 13-th column of the data line.
\item
Items can be separated by an arbitrary numbers of comment lines.
A line is treated as a comment and ignored
provided it does not contain in columns 1 through 10 any of the
valid keywords.
\item\label{twoitem}
Two items contain no data lines. The first one is defined by the
keyword {\tt{}EXECUTE}, and starts a calculation for the currently
defined input parameters. The second one is defined by the
keyword {\tt{}ALL\_DONE}, and terminates the program. The latter item
is not required if the compiler is able to properly recognize
the end of input data file. When the end of input data file is
encountered, the program terminates as if the keyword {\tt{}ALL\_DONE}
was found.
\item
A given item may occur any number of times in different
places of the input data file. Only the last one of the same items
occurring before a given {\tt{}EXECUTE} keyword is
taken into account.
\item
Between two consecutive items {\tt{}EXECUTE}, the order of other
items is arbitrary.
These other items define the data set, i.e, the set of
input parameters, for which the calculation
will start at the moment when the next item {\tt{}EXECUTE} is found.
\item
All input parameters have the default values defined in the code
(subroutine \pr{PREDEF}). Therefore, if the input data file
contains only the item {\tt{}EXECUTE}, and no other lines,
a calculation for the default values will be performed.
In the present version, this will result in performing 50 iterations
for the superdeformed state of {\DY} at the angular
frequency of $\ho$=0.5\,MeV.
\item
Only those items which define the values of input parameters which are
different than the default values
have to be included in the input data file. The values of input parameters
defined by any item will stay in effect till another occurrence
of the same item, or till the code terminates. This rule is valid
irrespectively of how many {\tt{}EXECUTE} items follow the given item.
\end{enumerate}

Together with the FORTRAN source code in the file {\tt{}hfodd.f},
two input data files
are provided, {\tt{}dy152-a.dat} and {\tt{}dy152-b.dat}.
The first one contains only one line with the keyword
{\tt{}EXECUTE}. The second one, which is reproduced in the
section TEST RUN INPUT below, contains all valid keywords
and the input parameters identical to the default values.
Therefore, the results of executing the program for the both
provided input data files are strictly identical. The second
input data file serves only as a suitable pattern to modify
the input parameters. However, it should be a good practice to
include in the input data file only those items which
modify the input parameters with respect to the default values.
In this way, a danger of accidentally modifying some intricate
numerical parameters of the code is minimized.

In the following subsections we discuss all the valid keywords
and their corresponding data lines. The default values are
given in the data lines following the keyword lines. In the data
lines we also show names of the corresponding variables
which are referred to in the text.
The meaning of several parameters that are introduced below in the context
of their function in the code is explained in detail in I.

\subsection{General data}
\label{sec6o}

\key{NUCLID} 86, 66 = {\tt{}IN\_FIX}, {\tt{}IZ\_FIX}

\noindent Numbers of neutrons ({\tt{}IN\_FIX}) and protons ({\tt{}IZ\_FIX})
          in the nucleus under consideration. Calculations for odd
          and odd-odd nuclei require {\tt{}IROTAT}=1.

\key{ITERATIONS} 50 = {\tt{}NOITER}

\noindent For {\tt{}NOITER}$>$0, the specified number of iterations
          is performed. The code starts
          with the iteration number 0 by constructing the initial
          potentials. Specifying {\tt{}NOITER}=0 requests
          only this initial phase. If the iteration is restarted from
          the previously recorded potentials ({\tt{}ICONTI}=1), then the
          counting of iterations continues from the previous value on.

\subsection{Interaction}
\label{sec6l}

\key{SKYRME-SET} SKM* = {\tt{}SKYRME}

\noindent {\tt{}CHARACTER*4} acronym of the Skyrme force parameter set.
          Must start at the 13-th column of the data line.
          At present, valid acronyms are {SKM*}, {SIII}, {SKP},
          and {SKI1}.
          Other sets of parameters can easily be included in the
          subroutine \pr{PARAMS}.

\vspace{1ex}
\noindent\begin{tabular}{@{}ll@{}l@{}l}
{\it Keyword:}{\spc}{\tt{}EVE\_SCA\_TS}&&&\\
             & \multicolumn{3}{l}{1., 1., 1., 1., 1., 1.,
                                  1., 1., 1., 1., 1., 1.} \\
             & {\tt{}SRHO\_T}, {\tt{}SRHO\_S,}& {\tt{}SRHODT}, {\tt{}SRHODS,}
             & {\tt{}SLPR\_T}, {\tt{}SLPR\_S,}                             \\
           &&& {\tt{}STAU\_T}, {\tt{}STAU\_S,}                             \\
           &&& {\tt{}SSCU\_T}, {\tt{}SSCU\_S,}                             \\
           &&& {\tt{}SDIV\_T}, {\tt{}SDIV\_S }
\end{tabular}

\noindent By using this item in the data set
          the coupling constants corresponding to a given Skyrme parameter
          set can be arbitrarily scaled. This allows calculations with
          modified Skyrme functionals \cite{J2D2b}.
          The time-even coupling constants in the total-sum representation
          (I-\ref{eq582}) are multiplied by the 12 numbers from
          {\tt{}SRHO\_T} to {\tt{}SDIV\_S}.
          The variables with names ending with {\tt{}\_T} and
          {\tt{}\_S} multiply the ``total'' and ``sum'' coupling constants,
          respectively. The variables with names containing the
          acronyms {\tt{}RHO}, {\tt{}LPR}, {\tt{}TAU}, {\tt{}SCU}, and
          {\tt{}DIV}, multiply the coupling constants with superscripts
          $\rho$, $\Delta\rho$, $\tau$, $J$, and $\nabla{J}$, respectively,
          and those with {\tt{}RHOD} multiply the density-dependent part
          of $C^\rho$. Similar name convention is used for many other
          variables in the code \pr{HFODD}.

\vspace{1ex}
\noindent\begin{tabular}{@{}ll@{}l@{}l}
{\it Keyword:}{\spc}{\tt{}ODD\_SCA\_TS}&&&\\
             & \multicolumn{3}{l}{1., 1., 1., 1., 1., 1.,
                                  1., 1., 1., 1., 1., 1.} \\
             & {\tt{}SSPI\_T}, {\tt{}SSPI\_S,}& {\tt{}SSPIDT}, {\tt{}SSPIDS,}
             & {\tt{}SLPS\_T}, {\tt{}SLPS\_S,}                             \\
           &&& {\tt{}SCUR\_T}, {\tt{}SCUR\_S,}                             \\
           &&& {\tt{}SKIS\_T}, {\tt{}SKIS\_S,}                             \\
           &&& {\tt{}SROT\_T}, {\tt{}SROT\_S }
\end{tabular}

\noindent Same as above but for the time-odd coupling constants. Acronyms
          {\tt{}SPI}, {\tt{}LPS}, {\tt{}CUR}, {\tt{}KIS}, and {\tt{}ROT}
          correspond to coupling constants with superscripts
          $s$, $\Delta{s}$, $T$, $j$, and $\nabla{j}$, respectively,
          and those with {\tt{}SPID} correspond to the density-dependent part
          of $C^s$.

\vspace{1ex}
\noindent\begin{tabular}{@{}ll@{}l@{}l}
{\it Keyword:}{\spc}{\tt{}EVE\_SCA\_PM}&&&\\
             & \multicolumn{3}{l}{1., 1., 1., 1., 1., 1.,
                                  1., 1., 1., 1., 1., 1.} \\
             & {\tt{}SRHO\_P}, {\tt{}SRHO\_M,}& {\tt{}SRHODP}, {\tt{}SRHODM,}
             & {\tt{}SLPR\_P}, {\tt{}SLPR\_M,}                             \\
           &&& {\tt{}STAU\_P}, {\tt{}STAU\_M,}                             \\
           &&& {\tt{}SSCU\_P}, {\tt{}SSCU\_M,}                             \\
           &&& {\tt{}SDIV\_P}, {\tt{}SDIV\_M }
\end{tabular}

\noindent Same as above but for the time-even coupling constants in the
          isoscalar-isovector representation, Eq.{\spc}(I-\ref{eq581}).
          The variables with names ending with {\tt{}\_P} and
          {\tt{}\_M} multiply the isoscalar and isovector coupling constants,
          respectively.
          The total-sum scaling factors are used first, and the
          isoscalar-isovector scaling factors are used afterwards.

\vspace{1ex}
\noindent\begin{tabular}{@{}ll@{}l@{}l}
{\it Keyword:}{\spc}{\tt{}ODD\_SCA\_PM}&&&\\
             & \multicolumn{3}{l}{1., 1., 1., 1., 1., 1.,
                                  1., 1., 1., 1., 1., 1.} \\
             & {\tt{}SSPI\_P}, {\tt{}SSPI\_M,}& {\tt{}SSPIDP}, {\tt{}SSPIDM,}
             & {\tt{}SLPS\_P}, {\tt{}SLPS\_M,}                             \\
           &&& {\tt{}SCUR\_P}, {\tt{}SCUR\_M,}                             \\
           &&& {\tt{}SKIS\_P}, {\tt{}SKIS\_M,}                             \\
           &&& {\tt{}SROT\_P}, {\tt{}SROT\_M }
\end{tabular}

\noindent Same as above but for the time-odd coupling constants in the
          isoscalar-isovector representation, Eq.{\spc}(I-\ref{eq581}).

\key{G\_SCALING} 1.0, 1.0 = {\tt{}FACTGN}, {\tt{}FACTGP}

\noindent For {\tt{}IPAIRI}=1 the code \pr{HFODD} solves the
          BCS equations with the neutron and proton pairing
          strengths defined in Ref.{\spc}\cite{DMS80}.
          These values can be modified by defining here the
          multiplicative factors {\tt{}FACTGN} and {\tt{}FACTGP}
          for neutrons and protons, respectively.

\subsection{Symmetries}
\label{sec6f}

\key{ROTATION} 1  =  {\tt{}IROTAT}

\noindent Calculation with time-reversal breaking will be performed
          for {\tt{}IROTAT}=1, while the time-reversal symmetry
          will be conserved for {\tt{}IROTAT}=0. In the latter case the
          calculations will be performed only for one value of the
          simplex, $s$=+$i$, which gives almost twice shorter execution
          times. {\tt{}IROTAT}=0 is incompatible with providing
          a non-zero value of the angular frequency or with attempting
          a calculation for an odd or odd-odd nucleus.

\key{SIGNATUREY} 1  =  {\tt{}ISIGNY}

\noindent Calculation with parity/signature symmetry conserved
          are performed for {\tt{}ISIGNY}=1, while the parity
          (and signature) will be broken for {\tt{}ISIGNY}=0.
          In the former case the parity blocks will be separately
          diagonalized and the densities will be summed only
          in the 1/8 of all the Gauss-Hermite nodes. This may give
          almost twice shorter execution times. {\tt{}ISIGNY}=1
          requires {\tt{}ISIMTX}={\tt{}ISIMTZ}=1, and {\tt{}ISIGNY}=0
          requires {\tt{}ISIMTX}=0, or {\tt{}ISIMTZ}=0, or
          {\tt{}ISIMTX}={\tt{}ISIMTZ}=0, see Sec.{\spc}\ref{sec9d} of I.

\key{TSIMPLEXES} 1, 1 = {\tt{}ISIMTX}, {\tt{}ISIMTZ}

\noindent Calculation with conserved symmetries given by
          $x$-simplex$^T$ and/or $z$-simplex$^T$,
          see Sec.{\spc}\ref{sec9d} of I,
          will be performed for {\tt{}ISIMTX}=1 and/or {\tt{}ISIMTZ}=1,
          respectively.
          These symmetries will be broken for
          {\tt{}ISIMTX}=0 and {\tt{}ISIMTZ}=0.
          Values of {\tt{}ISIMTX} and {\tt{}ISIMTZ}
          should be compatible with
          that of {\tt{}ISIGNY} in such a way that
          {\tt{}ISIGNY}={\tt{}ISIMTX*ISIMTZ}.

\key{PAIRING} 0 = {\tt{}IPAIRI}

\noindent Calculation without pairing correlations (pure HF)
          will be performed for {\tt{}IPAIRI}=0 while the
          BCS pairing correlations will be included for
          {\tt{}IPAIRI}=1. The latter case is incompatible with
          {\tt{}IROTAT}=1.

\subsection{Configurations}
\label{sec6e}

\key{VACSIM\_NEU} 43, 43 = {\tt{}KVASIM(0)}, {\tt{}KVASIM(1)}

\noindent Numbers of lowest neutron states occupied in the two
          blocks, denoted by ($+$) and ($-$),
          of given simplexes, $s$=+$i$ and $s$=$-$$i$, respectively.
          These numbers define the simplex reference configuration from
          which the particle-hole excitations are counted.
          The definitions of simplex reference configuration and
          excitations are ignored unless {\tt{}ISIGNY}=0
          and {\tt{}IPAIRI}=0.

\key{VACSIM\_PRO} 33, 33 = {\tt{}KVASIM(0)}, {\tt{}KVASIM(1)}

\noindent Same as above but for the numbers of proton states.

\key{PHSIMP\_NEU} 1,   00, 00,   00, 00 = {\tt{}NUPAHO},
                                          {\tt{}KPSIMP}, {\tt{}KPSIMM},
                                          {\tt{}KHSIMP}, {\tt{}KHSIMM}

\noindent Neutron particle-hole excitations in the simplex blocks.
          {\tt{}NUPAHO} is the consecutive number from 1 to 5
          (up to five sets of excitations can be specified
          in separate items).
          Particles are removed from the {\tt{}KHSIMP}-th state
          in the ($+$) block and from the {\tt{}KHSIMM}-th state
          in the ($-$) block, and put in the {\tt{}KPSIMP}-th state
          in the ($+$) block and in the {\tt{}KPSIMM}-th state
          in the ($-$) block. At every stage of constructing
          excitations the Pauli exclusion principle has to be
          respected (particle removed from an occupied state
          and put in an empty state). Values equal zero have no effect.
          In practice, reasonable excitations can only be constructed
          by consulting the printed lists of single-particle
          states with their consecutive numbers in given blocks,
          see the TEST RUN OUTPUT below.

\key{PHSIMP\_PRO} 1,   00, 00,   00, 00 = {\tt{}NUPAHO},
                                          {\tt{}KPSIMP}, {\tt{}KPSIMM},
                                          {\tt{}KHSIMP}, {\tt{}KHSIMM}

\noindent Same as above but for the proton particle-hole excitations.

\vspace{1ex}
\noindent\begin{tabular}{@{}ll@{}l}
{\it Keyword:}{\spc}{\tt{}VACSIG\_NEU}&&\\
              & 22, 22, 21, 21 = &{\tt{}KVASIG(0,0)}, {\tt{}KVASIG(0,1)}, \\
                                &&{\tt{}KVASIG(1,0)}, {\tt{}KVASIG(1,1)}
\end{tabular}

\noindent Numbers of lowest neutron states occupied in the four
          blocks, denoted by ($+$,$+$), ($+$,$-$), ($-$,$+$), and ($-$,$-$),
          of given (parity,signature) combinations, i.e.,
          $(\pi,r)$=(+1,+$i$), (+1,$-$$i$), ($-$1,+$i$), and ($-$1,$-$$i$),
          respectively.
          These numbers define the parity/signature
          reference configuration from
          which the particle-hole excitations are counted.
          The definitions of parity/signature reference configuration and
          excitations are ignored unless {\tt{}ISIGNY}=1
          and {\tt{}IPAIRI}=0.

\vspace{1ex}
\noindent\begin{tabular}{@{}ll@{}l}
{\it Keyword:}{\spc}{\tt{}VACSIG\_PRO}&&\\
              & 16, 16, 17, 17 = &{\tt{}KVASIG(0,0)}, {\tt{}KVASIG(0,1)}, \\
                                &&{\tt{}KVASIG(1,0)}, {\tt{}KVASIG(1,1)}
\end{tabular}

\noindent Same as above but for the numbers of proton states.

\vspace{1ex}
\noindent\begin{tabular}{@{}ll@{}l}
{\it Keyword:}{\spc}{\tt{}PHSIGN\_NEU}&&\\
                    & 1, 00,00,00,00, 00,00,00,00 =&{\tt{}NUPAHO}, \\
                                     && {\tt{}KPPPSP}, {\tt{}KPPPSM},
                                        {\tt{}KPPMSP}, {\tt{}KPPMSM}, \\
                                     && {\tt{}KHPPSP}, {\tt{}KHPPSM},
                                        {\tt{}KHPMSP}, {\tt{}KHPMSM}
\end{tabular}

\noindent Neutron particle-hole excitations in the parity/signature blocks.
          Basic principles are the same as for the excitations
          in the simplex blocks.
          Particles are removed from the {\tt{}KHPPSP}-th state
          in the ($+$,$+$) block, from the {\tt{}KHPPSM}-th state
          in the ($+$,$-$) block, from the {\tt{}KHPMSP}-th state
          in the ($-$,$+$) block, and from the {\tt{}KHPMSM}-th state
          in the ($-$,$-$) block, and put in the {\tt{}KPPPSP}-th state
          in the ($+$,$+$) block, in the {\tt{}KPPPSM}-th state
          in the ($+$,$-$) block, in the {\tt{}KPPMSP}-th state
          in the ($-$,$+$) block, and in the {\tt{}KPPMSM}-th state
          in the ($-$,$-$) block.

\vspace{1ex}
\noindent\begin{tabular}{@{}ll@{}l}
{\it Keyword:}{\spc}{\tt{}PHSIGN\_PRO}&&\\
                    & 1, 00,00,00,00, 00,00,00,00 = &{\tt{}NUPAHO}, \\
                                     && {\tt{}KPPPSP}, {\tt{}KPPPSM},
                                        {\tt{}KPPMSP}, {\tt{}KPPMSM}, \\
                                     && {\tt{}KHPPSP}, {\tt{}KHPPSM},
                                        {\tt{}KHPMSP}, {\tt{}KHPMSM}
\end{tabular}

\noindent Same as above but for the proton particle-hole excitations.

\subsection{Ensemble of specific parameters referred to as ``Numerical data''}
\label{sec6p}

\key{MAX\_MULTIP} 2, 4, 4 = {\tt{}NMUCON}, {\tt{}NMUCOU}, {\tt{}NMUPRI}

\noindent Maximum multipolarities $\lambda$ of multipole moments
          used in the code for the constraints, Eq.{\spc}(I-\ref{eq507}),
          surface term of the
          Coulomb field, Eq.{\spc}(I-\ref{eq570}), and printed on the output,
          respectively. Values not larger than
          $\lambda$=9 are currently allowed.
          In case of the conserved parity/signature, only even
          multipoles are used in the Coulomb field.

\key{COULOMB} 80, 79, 0.25 = {\tt{}NUMCOU}, {\tt{}NUMETA}, {\tt{}FURMAX}

\noindent {\tt{}NUMCOU} gives the number of points
          $N^{\text{Coul}}$, Eq.{\spc}(I-\ref{eq579}), used when
          summing up the Coulomb Green function.
          The dimensionless parameter $d$ defining the size of
          the Coulomb parallelepiped,
          Eqs.{\spc}(I-\ref{eq572}) and (I-\ref{eq567}),
          is given by {\tt{}NUMCOU*FURMAX}.
          {\tt{}NUMETA} gives the order of the Simpson integration
          of the solid harmonics on the faces of the parallelepiped,
          see Sec.{\spc}\ref{sec4b} of I.

\key{SLOW-DOWN} 0.5, 0.5 = {\tt{}SLOWEV}, {\tt{}SLOWOD}

\noindent The standard prescription to calculate the HF potential
          in the next iteration is to mix a given fraction $\epsilon$
          of the HF potentials from the previous iteration,
          with the fraction 1$-$$\epsilon$ of potentials
          given by expressions (I-\ref{eq209}). {\tt{}SLOWEV} and {\tt{}SLOWOD}
          give the values of $\epsilon$ separately for the time-even and
          time-odd potentials.

\key{EPS\_HERMIT} 1.0E--14 = {\tt{}EPSHER}

\noindent Numerical precision requested for
          determining values of Hermite polynomials.

\key{OPTI\_GAUSS} 1 = {\tt{}IOPTGS}

\noindent For {\tt{}IOPTGS}=1 and {\tt{}IREAWS}=0,
          expression (I-\ref{eq565}) is used to calculate
          the order of the Gauss-Hermite integration, and the input
          parameters {\tt{}NXHERM}, {\tt{}NYHERM}, and {\tt{}NZHERM}
          are ignored.

\key{GAUSHERMIT} 18, 18, 32 = {\tt{}NXHERM}, {\tt{}NYHERM}, {\tt{}NZHERM}

          Orders $L_x$, $L_y$, and $L_z$ of the
          Gauss-Hermite integration in three Cartesian
          directions. Must be even.
          Ignored if {\tt{}IOPTGS}=0 or {\tt{}IREAWS}=1.

\subsection{Parameters of the HO basis}
\label{sec6t}

\key{BASIS\_SIZE} 15, 301, 800.0 = {\tt{}NOSCIL}, {\tt{}NLIMIT}, {\tt{}ENECUT}

\noindent The HO basis is composed of states having not more than
          $N_0$={\tt{}NOSCIL} quanta in either of the Cartesian directions,
          and not more than $M$={\tt{}NLIMIT} states in total,
          see Sec.{\spc}\ref{sec5a}. The states
          are added to the basis
          according to the increasing energy of the
          deformed harmonic oscillator (\ref{eq592}). In case of degenerate
          HO states (e.g., for an axially deformed HO) the complete multiplets
          are included, so the actual number of states {\tt{}LDBASE}
          can be slightly larger than {\tt{}NLIMIT}.
          If {\tt{}NLIMIT}$>$0, the cut-off energy {\tt{}ENECUT}
          is ignored; otherwise all states having HO energy smaller
          then {\tt{}ENECUT} are included in the basis, and {\tt{}NLIMIT}
          is ignored.

\key{HOMEGAZERO} 1.2 = {\tt{}FCHOM0}

\noindent The code uses the standard value of the spherical
          HO frequency $\hozero$=41\,MeV/$A^{1/3}$
          multiplied by the scaling factor $f$={\tt{}FCHOM0},
          see Eq.{\spc}(\ref{eq594}).

\key{SURFAC\_PAR} 86, 66, 1.23 =
                {\tt{}INNUMB}, {\tt{}IZNUMB}, {\tt{}R0PARM}

\noindent The code \pr{HFODD} calculates parameters of the HO basis,
          and the zero-iteration Nilsson potential, by
          defining the standard classical surface $\Sigma$,
          Eq.{\spc}(\ref{eq101}), corresponding to the volume
          $\frac{4}{3}\pi${\tt{}*R0PARM}$^3${\tt{}*(INNUMB+IZNUMB)}.

\key{SURFAC\_DEF} 2, 0, 0.61 = {\tt{}LAMBDA}, {\tt{}MIU}, {\tt{}ALPHAR}

\noindent The code defines frequencies of the deformed HO
          in three directions by using relation (\ref{eq593}),
          $\omega_x{}R_x$=$\omega_y{}R_y$=$\omega_z{}R_z$,
          where $R_\mu$ are the
          lengths of principal axes of the nuclear surface $\Sigma$ defined
          in the standard way (\ref{eq101}) by real deformation parameters
          $\alpha_{\lambda\mu}$={\tt{}ALPHAR(LAMBDA,MIU)}.
          The overall factor is defined
          by $(\omega_x\omega_y\omega_z)^{1/3}$=$\omega_0$.
          For details see Sec.{\spc}\ref{sec5a}.

\subsection{Constraints}
\label{sec6r}

\key{OMEGAY} 0.5 = {\tt{}OMEGAY}

\noindent Value of the cranking rotational frequency
          $\omega_J$={\tt{}OMEGAY}, Eqs.{\spc}(I-\ref{eq508})
          and (I-\ref{eq512}). Non-zero
          value requires {\tt{}IROTAT}=1. Pure linear constraint
          on the value of spin, $\omega_y$=$\omega_J$
          is used for {\tt{}IFLAGI}=0.

\key{MULTCONSTR}  2, 0, 0.01, 42.0, 1  =
                  {\tt{}LAMBDA}, {\tt{}MIU},
                  {\tt{}STIFFQ}, {\tt{}QASKED}, {\tt{}IFLAGQ}

\noindent For {\tt{}IFLAGQ}=1, the mass multipole moment of the given
          multipolarity $\lambda$ and $\mu$ is constrained. Values of
                  {\tt{}LAMBDA}, {\tt{}MIU},
                  {\tt{}STIFFQ}, and {\tt{}QASKED} correspond respectively to
          $\lambda$, $\mu$, $C_{\lambda\mu}$,
          and $\bar{Q}_{\lambda\mu}$ in Eq.{\spc}(I-\ref{eq507}).
          For {\tt{}IFLAGQ}=0, there is no constraint in the given
          multipolarity.

\key{SPINCONSTR}  0.0, 0.0,  0 = {\tt{}STIFFI}, {\tt{}ASKEDI}, {\tt{}IFLAGI}

\noindent For {\tt{}IFLAGI}=1, the quadratic constraint on spin is
          used together with the linear constraint.
          Values of {\tt{}STIFFI} and {\tt{}ASKEDI}
                   correspond respectively to
          $C_J$ and $\bar{J}_y$ in Eq.{\spc}(I-\ref{eq508}).
          For {\tt{}IFLAGI}=0, there is no quadratic constraint on spin,
          (but there still can be the standard cranking linear constraint
          defined by {\tt{}OMEGAY}).

\subsection{Output-file parameters}
\label{sec6s}

\key{PRINT-ITER} 1, 0, 1 = {\tt{}IPRSTA}, {\tt{}IPRMID}, {\tt{}IPRSTO}

\noindent The code prints results for the first, middle, and/or last
          iteration if the corresponding parameters
          {\tt{}IPRSTA}, {\tt{}IPRMID}, and {\tt{}IPRSTO} equal 1.

\key{EALLMINMAX} --12.0, 0.0 = {\tt{}EMINAL}, {\tt{}EMAXAL}

\noindent The code prints tables of single-particle properties
          for states with values of the single-particle Routhians between
          {\tt{}EMINAL} and {\tt{}EMAXAL}. No table is printed
          unless {\tt{}EMINAL} $\leq$ {\tt{}EMAXAL}.

\subsection{Files}
\label{sec6k}

\key{REVIEWFILE} HFODD.REV = {\tt{}FILREV}

\noindent {\tt{}CHARACTER*68} file name of the review file.
          Must start at the 13-th column of the data line.
          The ASCII review file is written after calculating
          every data set (i.e., once per every {\tt{}EXECUTE}
          item), provided {\tt{}IREVIE}$>$0. The file is not rewound,
          so the results for several data sets can be accumulated
          in a single file. This is so provided the filename
          {\tt{}FILREV} is not changed between the {\tt{}EXECUTE}
          items.  The file contains all relevant
          parameters and results of calculation in a form suitable
          for reading by another program. It is ment as an interface
          to programs which analyze and/or plot the results.
          The file contains sections defined by
          keywords (different than keywords used in
          the input data file described here).
          The detailed structure of the review file is not documented
          in the present write up, and can be inferred from inspecting
          the specimen produced by the sample run,
          and from the subroutine \pr{REVIEW}.

\key{REVIEW} 2 = {\tt{}IREVIE}

\noindent The review file will not be written if {\tt{}IREVIE}=0.
          For {\tt{}IREVIE}=2,
          the table of single-particle properties is included
          in the review file
          in addition to other results which are written for
          {\tt{}IREVIE}=1.

\key{RECORDFILE} HFODD.REC = {\tt{}FILREC}

\noindent {\tt{}CHARACTER*68} file name of the record file.
          Must start at the 13-th column of the data line.
          The binary record file is written after each HF iteration.
          It contains complete information which allows restarting
          the iteration in another run of the code.
          To restart, one has to specify {\tt{}ICONTI}=1 and
          provide the name of the file by defining
          {\tt{}FILREP}.
          In case of the computer crush, the record file
          contains the results
          of the last performed iteration. Upon a successful
          completion of the given input data set it contains
          the results of the last performed iteration. The file
          is always rewound before it is written, so the results
          for consecutive iterations do not pile up.

\key{REPLAYFILE} HFODD.REP = {\tt{}FILREP}

\noindent {\tt{}CHARACTER*68} file name of the replay file.
          Must start at the 13-th column of the data line.
          The binary replay file with the name
          defined in {\tt{}FILREP} must exist if {\tt{}ICONTI}=1,
          and will be read. If the
          filenames {\tt{}FILREP} and {\tt{}FILREC} are identical,
          the replay file will be subsequently overwritten as a new
          record file. These feature is implemented to
          facilitate chaining of jobs which follow one another.

\key{WOODSAFILE} WOODS.WFN = {\tt{}FILWOO}

\noindent {\tt{}CHARACTER*68} file name of the Woods-Saxon file.
          Must start at the 13-th column of the data line.
          This file is read  provided {\tt{}IREAWS}=1.
          The binary Woods-Saxon file constitutes an interface between
          the Woods-Saxon code and the code \pr{HFODD}.
          It contains the Woods-Saxon wave functions
          and numerous other parameters which define
          the current calculation. {\em The parameters
          read from the Woods-Saxon file overwrite the values provided
          in the input data file}. Since the current version
          of the Woods-Saxon code will be published separately \cite{DDL},
          the feature of starting the iteration
          from the Woods-Saxon results is not documented in the
          present write up.

\subsection{Starting the iteration}
\label{sec6m}

\key{RESTART} 0 = {\tt{}ICONTI}

\noindent For {\tt{}ICONTI}=1, results stored in the replay file
          (written as a record file in a previous run)
          will be used to start the iteration. The replay
          file name should be provided by defining {\tt{}FILREP}.
          If the previous run was done with {\tt{}IREAWS}=1, the
          current run must also use {\tt{}IREAWS}=1, and the same
          Woods-Saxon file must be provided. This so because the
          Woods-Saxon file contains not only the information about
          the starting potential (which is ignored for {\tt{}ICONTI}=1)
          but also defines the HO basis.

\key{READ\_WOODS} 0 = {\tt{}IREAWS}

\noindent For {\tt{}IREAWS}=1, the results stored by the previously performed
          Woods-Saxon calculation
          will be used to start the iteration
          and to define the HO basis. For {\tt{}ICONTI}=1 and {\tt{}IREAWS}=1
          the Woods-Saxon file must also be provided, and will be used only
          to define the HO basis.

\vspace{1ex}
\noindent\begin{tabular}{@{}l@{}l}
{\it Keyword:}{\spc}{\tt{}NILSSONPAR}&\\
                  & 0, --1.175, --0.247,
                       --1.175, --0.352, 11.17, 11.17, 6.28 = \\
                  & {\tt{}NILDAT}, {\tt{}CNILSN}, {\tt{}DNILSN},
                                  {\tt{}CNILSP}, {\tt{}DNILSP}, \\
                  & {\tt{}HBANIX}, {\tt{}HBANIY}, {\tt{}HBANIZ}
\end{tabular}

\noindent For {\tt{}IREAWS}=0
          and {\tt{}ICONTI}=0, the code starts the calculation
          from the Nilsson potential. If {\tt{}NILDAT}=1, the Nilsson
          parameters $C$ and $D$
          [Ref.{\spc}\cite{RS80}, Eq.{\spc}(2.89)]
          are for neutrons given by {\tt{}CNILSN} and {\tt{}DNILSN}
          and for protons by {\tt{}CNILSP} and {\tt{}DNILSP}, while
          the HO deformation is defined by
          {\tt{}HBANIX}, {\tt{}HBANIY}, and {\tt{}HBANIZ}.
          If {\tt{}NILDAT}=0, the Nilsson
          parameters $C$ and $D$ are defined by Eq.{\spc}(2.91)
          and Table 2.3
          of Ref.{\spc}\cite{RS80}, while
          the HO deformation is the same as that of the HO basis defined
          in Sec.{\spc}\ref{sec6t}. In the latter case the Nilsson parameters
          read from the input file are ignored.


\section{Output file}
\label{sec6j}

Together with the FORTRAN source code in the file {\tt{}hfodd.f},
an example of the output file
is provided in {\tt{}dy152.out}.
Selected lines from this file are presented
in the section TEST RUN OUTPUT below. This output file corresponds to the
input file {\tt{}dy152-b.dat} reproduced in the section TEST RUN INPUT
below.  Most of the
information printed on the output file is self-explanatory.  Here
we only give some details which are not explicit in the output
file.

The output file begins with the information pertaining to the
general parameters of the calculation, then gives information
about the starting point of the iteration, provides the
convergence report, and finally contains the results calculated
at the last iteration.

Section {\tt{}CLASSICAL NUCLEAR SURFACE} lists the deformation
parameters used to define the nuclear surface,
Eq.{\spc}(\ref{eq101}), from which the basis parameters are
derived, as described in Sec.{\spc}\ref{sec5a}.  It also gives the
basis parameters such as the {\tt{}OSCILLATOR FREQUENCIES:}
{\tt{}HBAROX}, {\tt{}HBAROY}, and {\tt{}HBAROZ} corresponding to $\hox$,
$\hoy$, and $\hoz$.

Section {\tt{}PHYSICAL CONSTANTS} gives the values of $\hbar c$ in
MeV\,fm ({\tt{}H\_BARC}), of $\hbar c/e^2$ ({\tt{}HBCOE2}),
of the neutron and proton masses in MeV/$c^2$ ({\tt{}XMASSN} and
{\tt{}XMASSP}), of the kinetic-energy coefficient $\hbar^2/2m$ before
({\tt{}HBMASS}) and after the center-of-mass correction ({\tt{}HBMRPA}),
Eq.{\spc}(I-\ref{eq501}), of the elementary charge squared $e^2$ in
MeV\,fm ({\tt{}ECHAR2}), and of the coefficient preceding the
integral in the Coulomb exchange energy ({\tt{}COULEX}),
Eq.{\spc}(I-\ref{eq506}).  For details see the comments in the
\pr{SETBAS} subroutine.

Section {\tt{}OSCILLATOR LENGTHS}, {\tt{}CONSTANTS}, and
{\tt{}FREQUENCIES} gives the values of 1/$b_\mu$, $b_\mu$,
and $\ho_\mu$, respectively, which characterize the
HO basis in three Cartesian directions,
$\mu$=$x$, $y$, $z$.

Section {\tt{}BASIS CUT-OFF CONTROL PARAMETERS} gives maximum
numbers of the HO quanta in three directions {\tt{}NXMAXX},
{\tt{}NYMAXX}, and {\tt{}NZMAXX} corresponding to $N_x$, $N_y$, and
$N_z$, Eq.{\spc}(I-\ref{eq544}),
as well as the orders of the Gauss-Hermite quadratures
{\tt{}NXHERM}, {\tt{}NYHERM}, and {\tt{}NZHERM}, corresponding to $L_x$,
$L_y$, and $L_z$, Eq.{\spc}(I-\ref{eq563}).
It also gives the number $M$ of the HO states
included in the basis as requested in the input file
({\tt{}NLIMIT}) and as used in the calculation ({\tt{}LDBASE}).

Section {\tt{}SHAPE OF THE OSCILLATOR-BASIS DIAMOND} gives the
numbers of the HO quanta in a given direction for fixed numbers
of the HO quanta in both remaining directions.  The output is
arranged in such a way that the shape of the grid of points
$n_xn_yn_z$ is clearly visualized by projections in every of the
three directions.

Section {\tt{}PARAMETER SET} gives the name and the values of the
Skyrme force parameters, while the following section
{\tt{}COEFFICIENTS DEFINING THE SKYRME FUNCTIONAL} gives the
corresponding values of the coupling constants in the Skyrme
functional (I-\ref{eq109}).  These values take into account the
scaling factors, Sec.{\spc}\ref{sec6l}, which are printed in the
section {\tt{}SCALING FACTORS} unless all are equal 1.

Sections {\tt{}PARITY/SIGNATURE CONFIGURATIONS} or {\tt{}SIMPLEX
CONFIGURATIONS} give the vacuum and particle-hole configurations
requested in the input file for the case of the parity symmetry
conserved or not conserved, respectively.

Section {\tt{}CONVERGENCE REPORT} gives the list of performed
iterations.  For each iteration one line is printed with the
energy (I-\ref{eq504}), stability (I-\ref{eq599}), average values of
the quadrupole moments $Q_{20}$ and $Q_{22}$, total angular
momentum, angular frequency (I-\ref{eq512}), and the ratio of
energies ${\cal E}/\tilde{\cal E}$, see Sec.{\spc}\ref{sec2d} of I.
The line corresponding to
the last iteration is not printed because a more detailed
information is given for the final state below this section.

Section {\tt{}SINGLE-PARTICLE PROPERTIES} lists the
single-particle states calculated for the Nilsson, Woods-Saxon,
or Hartree-Fock Routhian operators.  For every state one line is printed
which gives the value of the single-particle Routhian,
cf.{\spc}Eq.{\spc}(\ref{eq516}), the consecutive numbers in the
parity/signature or simplex blocks, the quantum numbers
$[N,n_z,\Lambda]\Omega$ of the asymptotic Nilsson state
which has the largest component in the given state, the average
value of the parity operator (in \%), the average values of
projections of the intrinsic and total angular momenta (in $\hbar$), and
their ratio called the $g$-factor.

Sections {\tt{}MULTIPOLE MOMENTS} give the values of neutron,
proton, or mass multipole moments in units of (10\,fm)$^\lambda$.
Traditional normalization factors are used as explained in the
comments in the subroutine \pr{DEFUNI}.  In particular, the
$\lambda$=0 moment corresponds to the number of particles.

Sections {\tt{}ROOT-MEAN-SQUARE AND GEOMETRIC SIZES} gives the rms
average values of the radius and of the $x$, $y$, and $z$
coordinates.  In order to better visualize the size of the
nucleus, the geometric sizes are also calculated by multiplying the
rms radius by $\sqrt{5/3}$ and the rms coordinates by $\sqrt{5}$.

Section {\tt{}DENSITY INTEGRALS IN THE SKYRME FUNCTIONAL} gives
the integrals of products of densities, which appear in the
Skyrme functional (I-\ref{eq109}). The terms in the
functional are identified by the acronyms described in
Sec.{\spc}\ref{sec6l}.

Section {\tt{}CONTRIBUTIONS TO ENERGY IN THE SKYRME FUNCTIONAL}
gives the values of various terms which appear in the Skyrme
functional (I-\ref{eq109}).  These contributions are the
products of the coupling constants and of the density integrals
described above.  The sums of time-even and time-odd
contributions are also printed.

Section {\tt{}ANGULAR MOMENTA} gives the average values of the
total and intrinsic neutron, proton, and total angular momentum (in $\hbar$).
It also gives the corresponding values and contributions to the
first moment of inertia ${\cal J}^{(1)}$=$I/\omega$ (in $\hbar^2$/MeV).

Sections {\tt{}NEUTRON CONFIGURATIONS} and {\tt{}PROTON
CONFIGURATIONS} give a visual representation of states occupied
in the parity/signature or simplex blocks.  The lines denoted by
{\tt{}CONF:}  give the configurations requested in the input data
set, while those denoted by {\tt{}VACC:}  give the configurations
characterizing the given HF state.  By comparing the two sets one
can verify whether the requested configuration has been obtained,
and eventually devise a new configuration to be calculated.  The
consecutive numbers printed in the horizontal direction
correspond to the consecutive numbers in blocks printed in the
section {\tt{}SINGLE-PARTICLE PROPERTIES}.  By comparing the two
sections one can effectively associate the Nilsson labels with
the calculated configurations and also prepare the configuration
input data described in Sec.{\spc}\ref{sec6e}.

Section {\tt{}ENERGIES} gives a summary of the energies calculated
for the HF state.  The kinetic (I-\ref{eq501}), single-particle
(I-\ref{eq595}), and pairing (I-\ref{eq510}) energies are printed for
neutrons {\tt{}(NEU)}, protons {\tt{}(PRO)}, and all particles
{\tt{}(TOT)}.  The direct {\tt{}(DIR)} and exchange {\tt{}(EXC)} Coulomb
energies, Eqs.{\spc}(I-\ref{eq505a}) and (I-\ref{eq506}), are printed
together with their sum {\tt{}(TOT)}.  The multipole {\tt{}(MULT)}
and cranking {\tt{}(SPIN)} constraint energies,
Eqs.{\spc}(I-\ref{eq507}) and (I-\ref{eq508}), are printed together
with the corresponding corrections {\tt{}(CORR.)} given by
Eqs.{\spc}(I-\ref{eq600}) and (I-\ref{eq598}), respectively.  Then
the rearrangement energy (I-\ref{eq597}) is printed followed by the
value of the Routhian (I-\ref{eq500}) and the
spin-orbit and Skyrme energies, the latter two split in the time-even
{\tt{}(EVE)} and time-odd {\tt{}(ODD)} contributions.  Finally, the
total energies $\tilde{\cal E}$ (I-\ref{eq596}) and ${\cal E}$
(I-\ref{eq504}) are printed as {\tt{}(SP)} and {\tt{}(FUN)},
respectively, while their difference (I-\ref{eq599}) is printed as
the stability {\tt{}(STAB}).

\section{FORTRAN source file}
\label{sec6v}

The FORTRAN source code in is provided in the file {\tt{}hfodd.f}
and can be modified in several places which are described in this section.

\subsection{Dimensions of arrays}
\label{sec6d}

The code \pr{HFODD} uses the arrays' dimensions declared through the
{\tt{}PARAMETER} statements. This allows changing the dimensions
and adapting the size of the reserved memory to the problem being
solved. Whenever too small a dimension is defined the code aborts
with a message indicating the dimension which should be increased.
Substantial amount of memory is required only for arrays depending
on the following {\tt{}PARAMETER} values:

\vspace{1ex}
\noindent{\tt{}PARAMETER (NDMAIN=16)}

\noindent Should be larger or equal to the input
          parameter {\tt{}NOSCIL} defined in Sec.{\spc}\ref{sec6t}.

\vspace{1ex}
\noindent{\tt{}PARAMETER (NDBASE=307)}

\noindent Should be larger or equal to the input
          parameter {\tt{}NLIMIT} defined in Sec.{\spc}\ref{sec6t}.
          It should also be larger or equal to the actual
          size of the HO basis {\tt{}LDBASE}, which can be larger
          than {\tt{}NLIMIT} in case of degenerate HO states.

\vspace{1ex}
\noindent{\tt{}PARAMETER (NDXHRM=19,NDYHRM=19,NDZHRM=33)}

\noindent Should be respectively larger or equal to the input
          parameters {\tt{}NXHERM}, {\tt{}NYHERM}, and {\tt{}NZHERM}
          defined in Sec.{\spc}\ref{sec6p}.

On vector machines, parameters
{\tt{}NDXHRM}, {\tt{}NDYHRM}, {\tt{}NDZHRM}, and {\tt{}NDBASE}
should be odd
in order to minimize the risk of bank memory conflicts.
By the same token, parameter {\tt{}NDMAIN} should be even,
because it defines the matrix dimensions beginning with 0.

\subsection{Vectorization properties}
\label{sec6h}

As discussed in Sec{\spc}\ref{sec3d} of I, the code \pr{HFODD}
has to operate by using seven-fold nested short loops, and
this part does not perform well in a vector processor.
However, it turns out that the loops can be artificially made longer
in such a way that the final CPU time in a vector processor actually becomes
much
shorter. All the places where this trick has been applied can be identified
in the source file
by finding the lines beginning with {\tt{}CVECTOR} and {\tt{}CSCALAR},
for example:

\begin{verbatim}
CVECTOR
                        DO KZ=0,LAZOXY(NX,NY)+LAZOXY(MX,MY)
CSCALAR                 DO KZ=NZ+MZ,0,-2
                           RESULT=RESULT+COEF00(KZ,NZ,MZ,3)*T_AUXI(KZ)
                        END DO
\end{verbatim}

The line beginning with  {\tt{}CSCALAR} should be made active
on a scalar or superscalar machine. The line immediately below the line
which begins with {\tt{}CVECTOR} should be active on a vector
machine. The results of calculation do not depend on which
version of loops is activated.

\subsection{Library subroutines}
\label{sec6i}

The code \pr{HFODD} requires an external subroutine which diagonalizes
complex hermitian matrices. In the present version, the code calls
the NAGLIB subroutine  \pr{F02AXE}. The call to this subroutine
is performed through the interface subroutine \pr{ZHPEV}, which
can be replaced by the ESSL subroutine of the same name and the same
calling parameters.

In the present version, the code also uses an external subroutine \pr{CGEMM}
from the BLAS CRAY library, which multiplies three complex matrices.
By using the value of {\tt{}PARAMETER (I\_CRAY=0)} the call is diverted
to the functionally equivalent subroutine \pr{ZGEMUL}, which is
provided in the source file. However, the latter one can also
be replaced by the ESSL subroutine of the same name and the same
calling parameters.


\section{Acknowledgments}
\label{sec8a}

\bigskip
Useful comments by G.~Hackman,  P.-H.~Heenen,
and W.~Satu{\l}a are
gratefully acknowledged.  We thank P.-H.~Heenen for providing us
with results of test calculations performed with his HF code, and
K.~Burzy\'nski, W.D.~Luo, H.~Molique, and T.R.~Werner for help in
coding the subroutines \pr{SKFILD}, \pr{SOLHAR}, \pr{PHASES}, and
\pr{CYLXYZ}.  We would like to express our thanks
to the {\it Institut du D\'eveloppement et de Ressources en
Informatique Scientifique} (IDRIS) of CNRS, France, which
provided us with the computing facilities under Project
No.~960333.  This research was supported in part by the Polish
Committee for Scientific Research under Contract
No.~2~P03B~034~08, and by the computational grant from the
Interdisciplinary Centre for Mathematical and Computational
Modeling (ICM) of the Warsaw University.



\clearpage

{\bf\large TEST RUN INPUT}

{\tt\scriptsize
\baselineskip 1ex

\begin{verbatim}
                          ----------  General data  ----------
NUCLID
            86    66
ITERATIONS
            50
                          ----------  Interaction  -----------
SKYRME-SET
            SKM*
EVE_SCA_TS   RHO       RHOD      LPR       TAU       SCU       DIV
            1.  1.    1.  1.    1.  1.    1.  1.    1.  1.    1.  1.
ODD_SCA_TS   SPI       SPID      LPS       CUR       KIS       ROT
            1.  1.    1.  1.    1.  1.    1.  1.    1.  1.    1.  1.
EVE_SCA_PM   RHO       RHOD      LPR       TAU       SCU       DIV
            1.  1.    1.  1.    1.  1.    1.  1.    1.  1.    1.  1.
ODD_SCA_PM   SPI       SPID      LPS       CUR       KIS       ROT
            1.  1.    1.  1.    1.  1.    1.  1.    1.  1.    1.  1.
G_SCALING
            1.0  1.0
                          ----------  Symmetries  ------------
ROTATION
            1
SIGNATUREY
            1
TSIMPLEXES
            1      1
PAIRING
            0
                          ----------  Configurations  --------
VACSIM_NEU       SIMP  SIMM
                  43    43
VACSIM_PRO       SIMP  SIMM
                  33    33
PHSIMP_NEU          PART              HOLE
            1     00    00          00    00
PHSIMP_PRO          PART              HOLE
            1     00    00          00    00
VACSIG_NEU       PPSP  PPSM  PMSP  PMSM
                  22    22    21    21
VACSIG_PRO       PPSP  PPSM  PMSP  PMSM
                  16    16    17    17
PHSIGN_NEU             PARTICLES                     HOLES
            1     00    00    00    00        00    00    00    00
PHSIGN_PRO             PARTICLES                     HOLES
            1     00    00    00    00        00    00    00    00
                          ----------  Numerical data  --------
MAX_MULTIP
            2      4      4
COULOMB
            80    79   0.25
SLOW-DOWN
            0.5  0.5
EPS_HERMIT
            1.00E-14
OPTI_GAUSS
            1
GAUSHERMIT
            18    18    32
                          ---  Parameters of the HO basis  ---
BASIS_SIZE
            15   301   800.
HOMEGAZERO
            1.2
SURFAC_PAR
            86    66   1.23
SURFAC_DEF
            2      0   0.61
SURFAC_DEF
            4      0   0.10
                          ----------  Constraints  ----------
OMEGAY
            0.50
MULTCONSTR
            2      0   0.01   42.   1
SPINCONSTR
                       0.00    0.   0
                          ----  Output-file  parameters  ----
PRINT-ITER
            1      0      1
EALLMINMAX
            -12.   0.
                          -------------  Files  -------------
REVIEWFILE
            HFODD.REV
REVIEW
            2
RECORDFILE
            HFODD.REC
REPLAYFILE
            HFODD.REP
WOODSAFILE
            WOODS.WFN
                          -----  Starting the iteration  ----
RESTART
            0
READ_WOODS
            0
NILSSONPAR
            0   -1.175 -0.247   -1.175 -0.352    11.170  11.170  6.280
                          ----------  Calculate  ------------
EXECUTE
                          ----------  Terminate  ------------
ALL_DONE
\end{verbatim}
}

\clearpage

{\bf\large TEST RUN OUTPUT}

{\tt\scriptsize
\baselineskip 1ex

\begin{verbatim}
*******************************************************************************
*                                                                             *
*     HFODD    HFODD    HFODD    HFODD    HFODD    HFODD    HFODD    HFODD    *
*                                                                             *
*******************************************************************************
*                                                                             *
*                   SKYRME-HARTREE-FOCK CODE VERSION 1.60N                    *
*                                                                             *
*              ONE SYMMETRY-PLANE AND NO TIME-REVERSAL SYMMETRY               *
*                                                                             *
*                DEFORMED CARTESIAN HARMONIC-OSCILLATOR BASIS                 *
*                                                                             *
*******************************************************************************
*                                                                             *
*                     JACEK DOBACZEWSKI AND JERZY DUDEK                       *
*                                                                             *
*             CENTRE DE RECHERCHES NUCLEAIRES, STRASBOURG, 1993-96            *
*                                                                             *
*******************************************************************************

*******************************************************************************
*                                                                             *
*  CLASSICAL NUCLEAR SURFACE DEFINED FOR:                    N = 86   Z = 66  *
*                                                                             *
*******************************************************************************
*                                                                             *
*  AL10 =   ZERO  AL11 =   ZERO  .............  .............  .............  *
*                                                                             *
*  AL20 =  0.610  AL21 =   ZERO  AL22 =   ZERO  .............  .............  *
*                                                                             *
*  AL30 =   ZERO  AL31 =   ZERO  AL32 =   ZERO  AL33 =   ZERO  .............  *
*                                                                             *
*  AL40 =  0.100  AL41 =   ZERO  AL42 =   ZERO  AL43 =   ZERO  AL44 =   ZERO  *
*                                                                             *
*******************************************************************************
*                                                                             *
*                                           HOMEGA=  9.2190  FCHOM0=  1.2000  *
*                                                                             *
*  OSCILLATOR FREQUENCIES: HBAROX= 11.1998  HBAROY= 11.1998  HBAROZ=  6.2464  *
*                                                                             *
*  MOMENTS OF INERTIA:     XMOMFC= 90.2596  YMOMFC= 90.2596  ZMOMFC= 42.8300  *
*                                                                             *
*  CENTRES OF MASS:        CMSXFC=  0.0000  CMSYFC=  0.0000  CMSZFC=  0.0000  *
*                                                                             *
*******************************************************************************

*******************************************************************************
*                                                                             *
*   PHYSICAL CONSTANTS:      H_BARC=197.32891000      HBCOE2=137.03602000     *
*                                                                             *
*                            XMASSN=938.90590000      XMASSP=938.27231000     *
*                                                                             *
*                            HBMASS= 20.73620941      HBMRPA= 20.59978698     *
*                                                                             *
*                            ECHAR2=  1.43997841      COULEX= -1.06350868     *
*                                                                             *
*******************************************************************************
*                                                                             *
*   OSCILLATOR LENGTHS:      X= 1.9243099    Y= 1.9243099    Z= 2.5766958     *
*                                                                             *
*   OSCILLATOR CONSTANTS:    X= 0.5196668    Y= 0.5196668    Z= 0.3880939     *
*                                                                             *
*   OSCILLATOR FREQUENCIES:  X=11.1997759    Y=11.1997759    Z= 6.2464470     *
*                                                                             *
*******************************************************************************
*                                                                             *
*   BASIS CUT-OFF CONTROL PARAMETERS:  NXMAXX=   8  NYMAXX=   8  NZMAXX=  15  *
*                                                                             *
*   OPTIMUM NUMBERS OF  GAUSS POINTS:  NXHERM=  18  NYHERM=  18  NZHERM=  32  *
*                                                                             *
*                                      NLIMIT= 301  LDBASE= 306  MCOUNT=4096  *
*                                                                             *
*                                      ENECUT= 800.0000     ELIMIT= 112.7539  *
*                                                                             *
*******************************************************************************

*******************************************************************************
*                                                                             *
*   SHAPE OF THE OSCILLATOR-BASIS DIAMOND                                     *
*                                                                             *
*******************************************************************************
*                                                                             *
*        NZ ===>>>  0  1  2  3  4  5  6  7  8  9 10 11 12 13 14 15            *
*        MAX.NX =>  8  8  7  7  6  6  5  4  4  3  3  2  2  1  0  0            *
*                 ------------------------------------------------            *
*   NX= 0  (15)  |  8  8  7  7  6  6  5  4  4  3  3  2  2  1  0  0            *
*   NX= 1  (13)  |  7  7  6  6  5  5  4  3  3  2  2  1  1  0                  *
*   NX= 2  (12)  |  6  6  5  5  4  4  3  2  2  1  1  0  0                     *
*   NX= 3  (10)  |  5  5  4  4  3  3  2  1  1  0  0                           *
*   NX= 4  ( 8)  |  4  4  3  3  2  2  1  0  0                                 *
*   NX= 5  ( 6)  |  3  3  2  2  1  1  0                                       *
*   NX= 6  ( 5)  |  2  2  1  1  0  0                                          *
*   NX= 7  ( 3)  |  1  1  0  0                                                *
*   NX= 8  ( 1)  |  0  0                                                      *
*                                                                             *
*******************************************************************************

*******************************************************************************
*                                                                             *
* PARAMETER SET SKM*:  T0= -2645.00  T1=   410.00  T2=  -135.00  T3= 15595.00 *
*                                                                             *
* POWER=0.1667  W=130  X0=  0.09000  X1=  0.00000  X2=  0.00000  X3=  0.00000 *
*                                                                             *
*******************************************************************************
\end{verbatim}
\newpage
\begin{verbatim}
*******************************************************************************
*                                                                             *
*                 COEFFICIENTS DEFINING THE SKYRME FUNCTIONAL                 *
*                                                                             *
*******************************************************************************
*                                                                             *
*                 TOTAL(T)        SUM(S)        ISOSCALAR(P)  ISOVECTOR(M)    *
*                 --------        ------        ------------  ------------    *
*    CRHO_ =    -1382.012500    780.275000       -991.875000    390.137500    *
*    CRHOD =     1299.583333   -649.791667        974.687500   -324.895833    *
*    CLPR_ =      -85.312500     34.218750        -68.203125     17.109375    *
*    CTAU_ =       68.750000    -68.125000         34.687500    -34.062500    *
*    CSCU_ =        0.000000     68.125000         34.062500     34.062500    *
*    CDIV_ =      -65.000000    -65.000000        -97.500000    -32.500000    *
*                                                                             *
*    CSPI_ =      -59.512500    661.250000        271.112500    330.625000    *
*    CSPID =        0.000000   -649.791667       -324.895833   -324.895833    *
*    CLPS_ =        0.000000     34.218750         17.109375     17.109375    *
*    CCUR_ =      -68.750000     68.125000        -34.687500     34.062500    *
*    CKIS_ =        0.000000    -68.125000        -34.062500    -34.062500    *
*    CROT_ =      -65.000000    -65.000000        -97.500000    -32.500000    *
*                                                                             *
*******************************************************************************
*******************************************************************************
*                                                                             *
*  PARITY/SIGNATURE CONFIGURATIONS:                                           *
*                                                                             *
*                V A C U U M        P A R T I C L E S         H O L E S       *
*                ===========        =================         =========       *
*            (++) (+-) (-+) (--)   (++) (+-) (-+) (--)   (++) (+-) (-+) (--)  *
*                                                                             *
*  NEUTRONS:  22   22   21   21      0    0    0    0      0    0    0    0   *
*  PROTONS :  16   16   17   17      0    0    0    0      0    0    0    0   *
*                                                                             *
*******************************************************************************
*******************************************************************************
*                                                                             *
* CONVERGENCE REPORT                                                          *
*                                                                             *
*******************************************************************************
*                                                                             *
*  ITER     ENERGY     STABILITY      Q20     Q22     SPIN   OMEGA  HOW NICE  *
*                                                                             *
*    0   -559.936804 -759.775974    54.178  -0.027   98.777  0.500  0.424287  *
*    1  -1163.385944   88.001030    48.812   0.025   53.243  0.500  1.081832  *
*   48  -1208.769024    0.000468    41.807   0.067   49.581  0.500  1.000000  *
*   49  -1208.768731    0.000400    41.806   0.067   49.581  0.500  1.000000  *
*                                                                             *
*******************************************************************************
*******************************************************************************
*                                                                             *
*  SINGLE-PARTICLE PROPERTIES: HARTREE-FOCK                        NEUTRONS   *
*                                                                             *
*******************************************************************************
*                                                                             *
*  NO)   ENERGY (++,+-,-+,--) | N,nz,/\,OMEG>   <P>     JY      SY    GFACT   *
*                                                                             *
*  76)  -11.743 ( 0, 0,21, 0) | 5, 3, 2, 3/2>  -100   0.103  -0.092  -0.896   *
*  77)  -11.728 ( 0, 0, 0,20) | 5, 3, 2, 3/2>  -100   0.209  -0.101  -0.485   *
*  78)  -11.433 (19, 0, 0, 0) | 4, 1, 1, 1/2>   100   0.007  -0.182 -3.E+01   *
*  79)  -11.334 ( 0,19, 0, 0) | 4, 1, 3, 5/2>   100  -0.129  -0.040   0.307   *
*  80)  -11.241 (20, 0, 0, 0) | 4, 1, 3, 5/2>   100  -0.100  -0.189   1.885   *
*  81)  -11.137 (21, 0, 0, 0) | 6, 5, 1, 1/2>   100   1.119  -0.176  -0.157   *
*  82)  -11.069 ( 0,20, 0, 0) | 6, 5, 1, 1/2>   100   0.984   0.084   0.086   *
*  83)  -10.881 ( 0,21, 0, 0) | 4, 1, 1, 1/2>   100  -0.112  -0.201   1.784   *
*  84)  -10.379 (22, 0, 0, 0) | 6, 4, 2, 5/2>   100  -0.155  -0.064   0.414   *
*  85)  -10.365 ( 0,22, 0, 0) | 6, 4, 2, 5/2>   100   0.007  -0.024  -3.300   *
*  86)   -9.557 ( 0, 0, 0,21) | 7, 6, 1, 3/2>  -100   2.517   0.041   0.016   *
*  87)   -7.880 ( 0, 0,22, 0) | 5, 2, 1, 3/2>  -100   0.903   0.201   0.222   *
*  88)   -7.840 ( 0, 0, 0,22) | 5, 2, 1, 3/2>  -100   0.442   0.217   0.490   *
*  89)   -7.738 (23, 0, 0, 0) | 4, 0, 2, 5/2>   100  -0.247   0.161  -0.650   *
*  90)   -7.732 ( 0,23, 0, 0) | 4, 0, 2, 5/2>   100  -0.235   0.162  -0.690   *
*  91)   -7.562 ( 0, 0,23, 0) | 5, 2, 1, 3/2>  -100   1.068   0.153   0.144   *
*  92)   -7.221 ( 0, 0, 0,23) | 5, 1, 4, 9/2>  -100  -0.337   0.028  -0.083   *
*  93)   -7.221 ( 0, 0,24, 0) | 5, 1, 4, 9/2>  -100  -0.337   0.028  -0.083   *
*  94)   -7.096 ( 0, 0, 0,24) | 7, 7, 0, 1/2>  -100  -0.134  -0.167   1.252   *
*  95)   -7.076 ( 0,24, 0, 0) | 6, 4, 0, 1/2>   100   1.232   0.328   0.266   *
*  96)   -6.350 ( 0,25, 0, 0) | 4, 0, 0, 1/2>   100   0.078   0.455   5.825   *
*  97)   -6.284 (24, 0, 0, 0) | 6, 3, 3, 7/2>   100   0.140  -0.044  -0.313   *
*                                                                             *
*******************************************************************************
*******************************************************************************
*                                                                             *
*                  DENSITY INTEGRALS IN THE SKYRME FUNCTIONAL                 *
*                                                                             *
*******************************************************************************
*                                                                             *
*                   TOTAL(T)        SUM(S)      ISOSCALAR(P)  ISOVECTOR(M)    *
*                   --------        ------      ------------  ------------    *
*    DRHO_ =       17.566618      8.912767         17.566618      0.258917    *
*    DRHOD =       12.488132      6.334617         12.488132      0.181102    *
*    DLPR_ =       -3.719401     -1.886188         -3.719401     -0.052976    *
*    DTAU_ =       15.598603      7.983209         15.598603      0.367815    *
*    DSCU_ =        0.119929      0.064074          0.119929      0.008219    *
*    DDIV_ =        0.823496      0.419204          0.823496      0.014912    *
*                                                                             *
*    DSPI_ =        0.018303      0.010522          0.018303      0.002740    *
*    DSPID =        0.013147      0.007544          0.013147      0.001941    *
*    DLPS_ =       -0.037372     -0.023155         -0.037372     -0.008937    *
*    DCUR_ =        0.060724      0.031580          0.060724      0.002437    *
*    DKIS_ =        0.029460      0.015277          0.029460      0.001094    *
*    DROT_ =        0.007844      0.004322          0.007844      0.000799    *
*                                                                             *
*******************************************************************************
\end{verbatim}
\newpage
\begin{verbatim}
*******************************************************************************
*                                                                             *
*               CONTRIBUTIONS TO ENERGY IN THE SKYRME FUNCTIONAL              *
*                                                                             *
*******************************************************************************
*                                                                             *
*                 TOTAL(T)        SUM(S)        ISOSCALAR(P)  ISOVECTOR(M)    *
*                 --------        ------        ------------  ------------    *
*    ERHO_ =   -24277.285107   6954.409417     -17423.888833    101.013143    *
*    ERHOD =    16229.367796  -4116.181326      12172.025847    -58.839377    *
*    ELPR_ =      317.311377    -64.543010        253.674755     -0.906388    *
*    ETAU_ =     1072.403924   -543.856103        541.076525    -12.528704    *
*    ESCU_ =        0.000000      4.365047          4.085077      0.279970    *
*    EDIV_ =      -53.527218    -27.248236        -80.290827     -0.484627    *
*               ============  ============      ============  ============    *
*    SUM EVEN:  -6711.729227   2206.945789      -4533.317455     28.534017    *
*                                                                             *
*    ESPI_ =       -1.089276      6.957479          4.962258      0.905945    *
*    ESPID =        0.000000     -4.901833         -4.271249     -0.630584    *
*    ELPS_ =        0.000000     -0.792320         -0.639413     -0.152907    *
*    ECUR_ =       -4.174748      2.151393         -2.106350      0.082995    *
*    EKIS_ =        0.000000     -1.040747         -1.003485     -0.037261    *
*    EROT_ =       -0.509892     -0.280907         -0.764838     -0.025961    *
*               ============  ============      ============  ============    *
*    SUM  ODD:     -5.773917      2.093065         -3.823078      0.142226    *
*                                                                             *
*******************************************************************************
*******************************************************************************
*                                                                             *
*  MULTIPOLE MOMENTS IN UNITS OF (10 FERMI)**LAMBDA                   TOTAL   *
*                                                                             *
*******************************************************************************
*                                                                             *
*  Q00 =152.0000  .............  .............  .............  .............  *
*                                                                             *
*  Q10 =    ZERO  Q11 =    ZERO  .............  .............  .............  *
*                                                                             *
*  Q20 = 41.8059  Q21 =    ZERO  Q22 =  0.0675  .............  .............  *
*                                                                             *
*  Q30 =    ZERO  Q31 =    ZERO  Q32 =    ZERO  Q33 =    ZERO  .............  *
*                                                                             *
*  Q40 =  4.7902  Q41 =    ZERO  Q42 =  0.0058  Q43 =    ZERO  Q44 = -0.0011  *
*                                                                             *
*******************************************************************************
*******************************************************************************
*                                                                             *
*  ROOT-MEAN-SQUARE AND GEOMETRIC SIZES IN FERMIS                     TOTAL   *
*                                                                             *
*  R_RMS =  5.5420    X_RMS =  2.3844    Y_RMS =  2.3712    Z_RMS =  4.4052   *
*                                                                             *
*  R_GEO =  7.1547    X_GEO =  5.3318    Y_GEO =  5.3022    Z_GEO =  9.8504   *
*                                                                             *
*******************************************************************************
*******************************************************************************
*                                                                             *
*  ANGULAR MOMENTA AND THE FIRST MOMENTS OF INERTIA FOR OMEGA = 0.500000 MEV  *
*                                                                             *
*******************************************************************************
*                                                                             *
*                            SPINS                            J(1)            *
*                ---------------------------     ---------------------------  *
*                ORBITAL   INTRINSIC   TOTAL     ORBITAL   INTRINSIC   TOTAL  *
*                                                                             *
*  NEUTRONS     28.21798   1.21125  29.42923    56.43596   2.42250  58.85846  *
*   PROTONS     19.29890   0.85368  20.15257    38.59779   1.70736  40.30515  *
*  --------                                                                   *
*     TOTAL     47.51687   2.06493  49.58181    95.03375   4.12986  99.16361  *
*                                                                             *
*******************************************************************************
*                                                                             *
*                           NEUTRON  CONFIGURATIONS                           *
*                           =======================                           *
*        P S  12 13 14 15 16 17 18 19 20 21 22 23 24 25 26 27 28 29 30 31 32  *
*        ---  --------------------------------------------------------------  *
*                                                                             *
*  CONF: + +   1  1  1  1  1  1  1  1  1  1  1  0  0  0  0  0  0  0  0  0  0  *
*  VACC: + +   1  1  1  1  1  1  1  1  1  1  1  0  0  0  0  0  0  0  0  0  0  *
*                                                                             *
*  CONF: + -   1  1  1  1  1  1  1  1  1  1  1  0  0  0  0  0  0  0  0  0  0  *
*  VACC: + -   1  1  1  1  1  1  1  1  1  1  1  0  0  0  0  0  0  0  0  0  0  *
*                                                                             *
*  CONF: - +   1  1  1  1  1  1  1  1  1  1  0  0  0  0  0  0  0  0  0  0  0  *
*  VACC: - +   1  1  1  1  1  1  1  1  1  1  0  0  0  0  0  0  0  0  0  0  0  *
*                                                                             *
*  CONF: - -   1  1  1  1  1  1  1  1  1  1  0  0  0  0  0  0  0  0  0  0  0  *
*  VACC: - -   1  1  1  1  1  1  1  1  1  1  0  0  0  0  0  0  0  0  0  0  0  *
*                                                                             *
*******************************************************************************
*                                                                             *
*                                ENERGIES (MEV)                               *
*                                                                             *
*******************************************************************************
*                                                                             *
*  KINETIC: (NEU)=  1652.730622    (PRO)=  1107.885480    (TOT)=  2760.616103 *
*  SUM EPS: (NEU)= -2034.973323    (PRO)= -1133.181733    (TOT)= -3168.155056 *
*  PAIRING: (NEU)=     0.000000    (PRO)=     0.000000    (TOT)=     0.000000 *
*                                                                             *
*  COULOMB: (DIR)=   564.434280    (EXC)=   -25.354572    (TOT)=   539.079708 *
*                                                                             *
*  CONSTR. (MULT)=     0.000377    SLOPE=    -0.003881    CORR.=    -0.081133 *
*  CONSTR. (SPIN)=   -24.790903    SLOPE=     0.500000    CORR.=   -12.395451 *
*                                                                             *
*  REARRANGEMENT ENERGY FROM THE SKYRME DENSITY-DEPENDENT TERMS=  1009.023720 *
*  ROUTHIAN  (TOTAL ENERGY PLUS MULTIPOLE AND SPIN CONSTRAINTS)= -1233.559005 *
*                                                                             *
*  SPIN-ORB (EVE)=   -80.775454    (ODD)=    -0.790799    (TOT)=   -81.566253 *
*  SKYRME:  (EVE)= -4504.783438    (ODD)=    -3.680852    (TOT)= -4508.464290 *
*                                                                             *
*  TOTAL:  (STAB)=     0.000343     (SP)= -1208.768136    (FUN)= -1208.768479 *
*                                                                             *
*******************************************************************************
\end{verbatim}
}


\clearpage

\begin{table}[ht]
\caption[T1]{%
Precision of results obtained for various observables calculated
by the code \pr{HFODD} with the physical basis parameters and
$M$=300.  Absolute values correspond to the results for {\Dy} and {\Pb},
while the relative values correspond to the differences between {\Tb}
and {\Dy}. The values denoted as ``precision'' represent also typical
orders of magnitude of the accuracy obtained in other cases for
the same numerical conditions.
}
\label{tab01}

\vspace{1ex}
\renewcommand{\arraystretch}{1.5}
\begin{center}
\begin{tabular}{|l|ll|ll|}
\hline
    &  \multicolumn{2}{c|}{Absolute values}  &
       \multicolumn{2}{c|}{Relative values} \\
\hline
Observable              &  Precision   &  \%    &  Precision    &  \%     \\
\hline
Energy ${\cal E}$       &      5\,MeV  &  0.4\% &     0.05\,MeV &  1\%    \\
Radius $R_{\text{rms}}$ &   0.003\,fm  & 0.05\% &        ---    &  ---    \\
Quadrupole moment $Q_p$ &     0.05\,b  &  0.3\% &       0.01\,b &  1\%    \\
Angular momentum  $I$   & 0.1\,$\hbar$ &  0.2\% & 0.03\,$\hbar$ &  2\%    \\
Static moment  $\jun$   &  0.2\,{\umi} &  0.2\% & 0.06\,{\umi}  &  2\%    \\
Dynamic moment $\jde$   &  0.1\,{\umi} &  0.1\% &   0.1\,{\umi} &  2-10\% \\
\hline
\end{tabular}
\end{center}
\end{table}

\clearpage

%
%

\begin{thebibliography}{99}

\bibitem{comcom1}
                 J. Dobaczewski and J. Dudek, Comput. Phys. Commun.,
                 the preceding paper.
\bibitem{Rei91}
                 P.-G. Reinhard, in {\it Computational Nuclear Physics I},
                 eds. K. Langanke, J.A. Maruhn, and S.E.Koonin
                 (Springer, Berlin, 1991) p. 28.
\bibitem{Gir83}
                 M. Girod and B. Grammaticos,
                 Phys. Rev. {\bf C27} (1983) 2317.
\bibitem{Bon85}
                 P. Bonche, H. Flocard, P.-H. Heenen, S.J. Krieger,
                 and M.S. Weiss,
                 Nucl. Phys. {\bf A443} (1985) 39.
\bibitem{Uma91}
                 A.S. Umar, M.R. Straayer, J.S. Wu, and M.C. G\"u{\c c}l\"u,
                 Phys. Rev. {\bf C44} (1991) 2512.
\bibitem{Chi95}
                 C.R. Chinn, A.S. Umar, M. Valli\'eres,
                 and M.R. Strayer,
                 Phys. Rev. {\bf E50} (1994) 5096;
                 Comput. Phys. Commun. {\bf 86} (1995) 40;
                 and references therein.
\bibitem{Dam69}
                 J. Damgaard, H.C. Pauli, V.V. Pashkevitch,
                 and V.M. Strutinsky, Nucl. Phys. {\bf A135} (1969) 432.
\bibitem{FQK73}
                 H. Flocard, P. Quentin, A.K. Kerman, and D. Vautherin,
                 Nucl. Phys. {\bf A203} (1973) 433.
\bibitem{SGN69}
                 S.G. Nilsson, C.-F. Tsang, A. Sobiczewski,
                 Z. Szyma\'nski, S. Wycech, C. Gustafson, I.-L.  Lamm,
                 P. M\"oller, and B. Nilsson,
                 Nucl. Phys. {\bf A131} (1969) 1.
\bibitem{DST81}
                 J. Dudek, Z. Szyma\'nski, and T.R. Werner,
                 Phys. Rev. {\bf C23} (1981) 920.
\bibitem{CDN87}
                 S. \'Cwiok, J. Dudek, W. Nazarewicz, J. Skalski,
                 and T. Werner, Comput. Phys. Commun. {\bf 46} (1987). 379
\bibitem{J2D2b}
                 J. Dobaczewski and J. Dudek,
                 Phys. Rev. {\bf C52} (1995) 1827.
\bibitem{BON87a}
                 P. Bonche, H. Flocard, and P.-H. Heenen,
                 Nucl. Phys. {\bf A467} (1987) 115.
\bibitem{BON91a}
                 P. Bonche, H. Flocard, and P.-H. Heenen,
                 Nucl. Phys. {\bf A523} (1991) 300.
\bibitem{PH}
                 P.-H. Heenen, private communication.
\bibitem{Bay86}
                 D. Baye and P.-H. Heenen, J. Phys. A {\bf 19} (1986) 2041.
\bibitem{Hee91}
                 P.-H. Heenen, P. Bonche, J. Dobaczewski,
                 H. Flocard, S.J. Krieger, J. Meyer, J. Skalski, N. Tajima,
                 and M.S. Weiss,
                 {\it International Workshop on Nuclear Structure Models},
                 Oak Ridge, 1992, eds. R. Bengtsson, J. Draayer, and
                 W. Nazarewicz (World Scientific, Singapore, 1992) p. 3.
\bibitem{Sat96}
                 W. Satu{\l}a, J. Dobaczewski, J. Dudek, and W. Nazarewicz,
                 Report nucl-th/9608019, to be published in Physical
                 Review Letters.
\bibitem{Bak95}
                 C. Baktash, B. Haas, and W. Nazarewicz, Annu. Rev. Nucl.
                 Part. Phys. {\bf 45} (1995) 485.
\bibitem{DMS80}
                 J. Dudek, A. Majhofer, and J. Skalski,
                 J. Phys. (London) {\bf G6}, 447 (1980).
\bibitem{DDL}
                 J. Dobaczewski, J. Dudek, and X. Li, to be published.
\bibitem{RS80}
                 P. Ring and P. Schuck, The Nuclear Many-Body Problem
                 (Springer, Berlin, 1980).
\end{thebibliography}
\end{document}